\documentclass[11pt]{article}
\pdfoutput=1 % if your are submitting a pdflatex (i.e. if you have
\usepackage{jheppub} % for details on the use of the package, please
\usepackage{xcolor}
\usepackage[T1]{fontenc} % if needed
\usepackage{amssymb}
\usepackage{amsmath}
\usepackage{color}

\newcommand{\Ima}{\text{Im}}

\newcommand{\Rea}{\text{Re}}
\newcommand{\Li}{\text{Li}_2}
\newcommand{\MZV}{\text{MZV}}
\newcommand{\pure}{\textrm{pure}}

\newcommand{\Disc}{\text{Disc}}
\newcommand{\dDisc}{\text{dDisc}}
\def\zb{\overline{z}}
\def\be{\begin{equation}}
\def\ee{\end{equation}}
\def\ba{\begin{equation}\begin{aligned}}
\def\ea{\end{aligned}\end{equation}}
\def\bea{\begin{eqnarray}}
\def\eea{\end{eqnarray}}
\def\cO{\mathcal{O}}
\def\cG{\mathcal{G}}
\def\P{C}
\def\Ppure{C^{\rm pure}}
\def\<{\langle}
\def\>{\rangle}
\def\Res{\operatorname*{Res}}
\def\kappat{\kappa'}

\title{Conformal Regge Theory At Finite Boost}
\author{Simon Caron-Huot}
\author{and Joshua Sandor}

\affiliation{Department of Physics, McGill University, 3600 Rue University, Montr\'eal, QC Canada}
\emailAdd{schuot@mcgill.ca}
\emailAdd{jsandor@stanford.edu}

\abstract{The Operator Product Expansion is a useful tool to represent correlation functions.
In this note we extend Conformal Regge theory to provide an exact OPE representation
of Lorenzian four-point correlators in conformal field theory, valid even away from Regge limit.
The representation extends convergence of the OPE by
rewriting it as a double integral over continuous spins and dimensions, and features a novel ``Regge block''.
We test the formula in the conformal fishnet theory, where exact results involving nontrivial Regge trajectories are available.}

\begin{document} 
\maketitle

\pagebreak
\section{Introduction}

The interactions between highly boosted objects is a topic of longstanding interest in relativistic field theory.
On the on hand, due to time dilation effects, the Regge limit (large boost with fixed impact parameter)
provides an instantaneous snapshot of essentially frozen objects.
On the other hand, since probes move near the lightcone,
observables in this limit are intrinsically dynamical and are strongly constrained by relativistic causality.

\smallskip

Many systems in the Regge limit exhibit a transient regime where interactions grow
as a function of boost, before saturating as required by quantum mechanical conservation of probability.
Regge theory quantifies the growth by the spin of effective excitations.
A famous example is the rising hadronic cross-sections attributed to so-called Pomeron exchanges.
Regge theory applies as well to highly boosted correlators in conformal field theories \cite{Brower:2006ea,Cornalba:2007fs,Costa:2012cb}.
In strongly coupled, holographic CFTs, the dominant effective excitation is nothing but the bulk
graviton. Its exchange grows as fast as allowed by the bound on chaos \cite{Maldacena:2015waa},
making the consistency constraints mentioned above particularly stringent.
Indeed, the fact that gravity grows with boost restricts its very structure at all energies \cite{Camanho:2014apa};
more generally, growing amplitudes must satisfy positivity properties related to the
Average Null Energy Condition \cite{Hartman:2016lgu}.

In many studies of the Regge limit,
it is often sufficient to consider only the leading term at large boost
(in the intermediate growth regime).
However, there may be situations where subleading effects are important.
A simple example would be to study effects from photons in addition to gravitons.
Another example would be saturation.
Finally there may be theories where interactions do \emph{not} grow, warranting precision studies.
It was recently argued that the critical three-dimensional O(N) and Ising models are of this type,
with Regge intercept less than unity ($j_*<1$).
This leads to transparent scattering at large boost \cite{Liu:2020tpf,Caron-Huot:2020ouj}.
In general, the Regge limit in conformal theories
probes intermediate operators of large scaling dimension \cite{Cornalba:2007zb,Li:2017lmh,Costa:2017twz}.
In transparent theories one might thus hope to use Regge theory, with exchange of a few dominant trajectories,
to precisely bound the heavy spectrum,
which could improve convergence of bootstrap calculations.

The goal of this paper is to extend formulas from Conformal Regge Theory
so as to retain the exact energy dependence of four-point correlators.
%, and can thus systematically account for subleading power corrections.

Generally, the OPE in a conformal field theory converges whenever two local operators act on the vacuum,
no matter where they are inserted  in spacetime (see \cite{Kravchuk:2020scc} for a review).
The physical picture of effective Reggeized particles however arises from the OPE
between an initial and a final state of a scattering process---often called the ``$t$-channel''---and
the OPE in such channels diverges.
As reviewed below, this divergence occurs after Euclidean correlation functions are analytically
continued to a ``Regge sheet''. Our main result is an exact resummation of the OPE which converges
on the Regge sheet.
Since correlators on the Euclidean sheet are well understood, we can state the result
in terms of a difference or discontinuity:
\ba
 \Disc_{14}\cG(z,\zb) = \!\!\!\!\!\int\limits_{-\frac{d-2}{2}-i\infty}^{-\frac{d-2}{2}+i\infty} \frac{dJ}{2\pi i}
\int\limits_{\frac{d}{2}-i\infty}^{\frac{d}{2}+i\infty} \frac{d\Delta}{2\pi i}\: \frac{e^{i\pi J}c^t(\Delta,J) + c^u(\Delta,J)}{\sin(\pi J)}
\frac{R^{(a,b)}_{\Delta,J}(z,\zb)}{2\kappa^{(a,b)}_{\Delta+J}} + \mbox{(subtractions)}\,. %^{\circlearrowleft},
\label{Regge disc intro}
\ea
The salient feature, familiar from (conformal) Regge theory \cite{Cornalba:2007fs,Costa:2012cb},
is that a discrete sum over spins has been replaced by an integral.
The power of Regge's idea is that this enlarges the radius of convergence of the OPE.
We expect eq.~\eqref{Regge disc intro} to converge anywhere on the Regge sheet.

The novel feature of eq.~\eqref{Regge disc intro}, in comparison with earlier work,
is the ``Regge block'' $R^{(a,b)}_{\Delta,J}(z,\zb)$, defined in eq.~\eqref{Rab} below,
which accounts for subleading power corrections.
Perhaps surprisingly, the Regge block is \emph{not} simply the conformal block that one
might have guessed from the leading-power formulas.
The Regge block can be defined as the unique solution to conformal Casimir equations
with a certain vanishing discontinuity. This combination turns out to cancel certain spurious poles,
and we find that it neatly packages terms which otherwise might have been split in other treatments.

Starting from eq.~\eqref{Regge disc intro}, concrete formulas for
order-by-order asymptotic expansions in a given model can be obtained, as detailed in eq.~\eqref{main}.

These formulas will be tested in the conformal fishnet model \cite{Gurdogan:2015csr}, a recently proposed limit of $\mathcal{N}=4$ SYM that retains only scalar fields, but remains integrable at the cost of sacrificing unitarity.
The OPE data corresponding to certain four-point correlators is known exactly and the correlators
can be expanded in the coupling in terms of known special functions (harmonic polylogarithms).
Using this expansion, we analytically continue correlators
to the Lorentzian regime and compare their high-energy behaviour with eq.~\eqref{Regge disc intro}.

This paper is organized as follows.  In section \ref{sec:prelim} we review kinematics of the Regge limit and
the required analytic continuation, and we review the fishnet model.
Section \ref{confregge} derives our exact formula for the Regge limit,
after reviewing analogous manipulations in the S-matrix context.
We also obtain a formula for the double-discontinuity and confirm that it inverts the
``Lorentzian inversion formula''.
In section \ref{fishnetsection} we test these formulas for various correlators in the fishnet model.
Section \ref{sec:conclusion} presents our brief conclusions.

\section{Preliminaries}
\label{sec:prelim}

\subsection{Review of conformal Regge kinematics \label{reggeReview}}
\begin{figure}[t]
	\def\svgwidth{0.7\linewidth}
	\centering{
		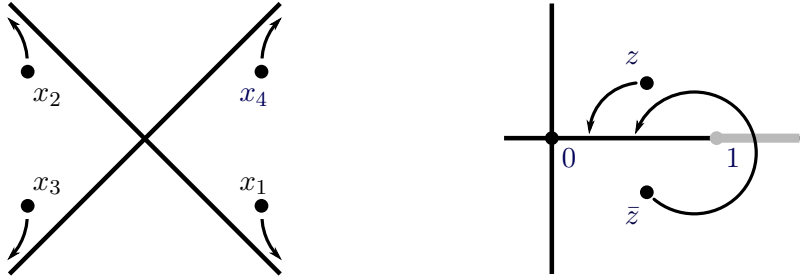
		\caption{Four-point kinematics in the CFT Regge regime. Panel (a) illustrates the kinematics, in which time runs vertically, and panel (b) illustrates the analytic continuation $\cG(z,\zb^\circlearrowleft)$: counterclockwise about the branch point. In the Euclidean regime $z^* = \zb$ but both $z$ and $\zb$ are real and independent in Lorentzian kinematics.
	\label{continuation}
	}}
\end{figure}

A conformal four-point correlator in Minkowski space $\mathcal{M}^{d-1,1}$ can be expressed as
\begin{equation}\label{corrfcn}
 \<\cO_4(x_4)\cO_3(x_3)\cO_2(x_2)\cO_1(x_1)\>=\frac{1}{(x_{12}^2)^{\frac{1}{2}(\Delta_1+\Delta_2)}(x_{34}^2)^{\frac{1}{2}(\Delta_3+\Delta_4)}} \bigg(\frac{x_{14}^2}{x_{24}^2}\bigg)^a\bigg(\frac{x_{14}^2}{x_{13}^2}\bigg)^b\cG(z,\zb),
\end{equation}where $a=\frac{1}{2}(\Delta_2-\Delta_1)$, $b=\frac{1}{2}(\Delta_3-\Delta_4)$ are combinations of the operators' scaling dimensions and the conformal cross-ratios $z$, $\zb$ are related to the coordinates $\{x_i\}$ by
\begin{equation} \label{zzb}
z\zb = \frac{(x_1-x_2)^2(x_3-x_4)^2}{(x_1-x_3)^2(x_2-x_4)^2}, \:\:\:\:\:\:\: (1-z)(1-\zb) = \frac{(x_1-x_4)^2(x_2-x_3)^2}{(x_1-x_3)^2(x_2-x_4)^2}.
\end{equation}
The Regge limit of the correlation function is attained by applying large and opposite boosts to the pairs (12) and (34), sendind
the operators to infinity along the lightcone:
\begin{equation}
x_1^+,x_3^- \to -\infty,\quad x_2^+,x_4^- \to +\infty.
\end{equation}
Here we have rewritten the vectors in lightcone coordinates $x_i = (x_i^+,x_i^-,x_i^a,x_i^b)$.
In the kinematics considered in this paper, both separations $(x_4-x_1)$ and $(x_2-x_3)$ are timelike while all other separations remain spacelike.
To evaluate the Regge limit, the Lorentzian correlator must be obtained from the Euclidean theory described above.
It is calculated by analytically continuing the theory from the region where $\zb=z^*$, namely by rotating $\zb$ around the branch point at $\zb=1$ while keeping $z$ fixed \cite{Costa:2012cb}.
The scattering process and analytic continuation are illustrated in figure \ref{continuation}.

To understand the continuation path a little more explicitly, we
recall that for Lorentzian correlators, time-like distances acquire a small imaginary part
$x_{23}^2\mapsto -|x_{23}|^2 \pm i0$ which is positive if the operators are in time-ordering
and negative otherwise.
The second cross-ratio in eq.~\eqref{zzb} thus accumulates
a phase $e^{2\pi i}$ which is indeed what happens along the path.

% JHEP will move figures to top anyway

By further defining $\sigma^2=z\zb$ and $w^2=z/\zb$, we see that the Regge limit corresponds to $\sigma\to0$ while $w$ is fixed. In analogy to the QFT Regge limit, we have an identification to the Mandelstam variables $\sigma \sim 1/s$ and $w \sim t$ in $s$-channel scattering.

In our chosen kinematics (with pairs $(1,4)$ and $(2,3)$ in separate Rindler wedges)
we have access to four operator orderings.
Two are equivalent, and rather trivial: if one pair is time-ordered and the other anti-time-ordered,
the continuation phases cancel out and the path does not leave the Euclidean sheet.
All novel Lorentzian information is contained in commutators, or discontinuities, of which we can define two natural ones:
\ba \label{Disc}
 \Disc_{23} \cG(z,\zb) &\equiv -ie^{i\pi (a+b)}\left(\cG(z,\zb^{\circlearrowleft})-\cG(z,\zb)\right)\,,\\
 \Disc_{14} \cG(z,\zb) &\equiv -i\left(e^{i\pi (a+b)}\cG(z,\zb^{\circlearrowleft}) - e^{-i\pi(a+b)}\cG(z,\zb)\right)\,.
\ea
The different phases originate from the prefactor in eq.~\eqref{corrfcn}.
These two discontinuities contain effectively the same information,
and the fourth independent operator ordering, $\cG(z,\zb^{\circlearrowright})$, can be reached by complex conjugation.

\subsection{Review of Conformal Fishnet Theory\label{fishnetreview}}

Conformal fishnet theory is a recently proposed integrable theory in $d=4$ that is neither a gauge theory nor supersymmetric \cite{Gurdogan:2015csr}.
A chief interest of this theory is the fact that very few Feynman diagrams contribute to any given process---often a unique diagram at each loop order
(or at each order in the `t Hooft $1/N$ expansion).
In this way, integrability of the theory allows for the calculation of certain Feynman diagrams which have been incalculable thus far by standard methods.

The theory contains two complex (matrix-valued) scalar fields,  and its simplification comes at the price of unitarity:
the basic 4-point interaction includes a term ${\rm Tr}(Y^\dagger X^\dagger YX)$ but not its complex conjugate.
Non-unitarity means that certain formulas below will contain unusual factors of the imaginary number $i$,
but there otherwise appears to be no obstructions to resumming perturbation theory and discuss finite-coupling correlators.

% C -> C/(4\pi)^4/(2\pi) = C/2/Pi^5

\begin{figure}[t]
	\centering{
		\def\svgwidth{0.7\linewidth}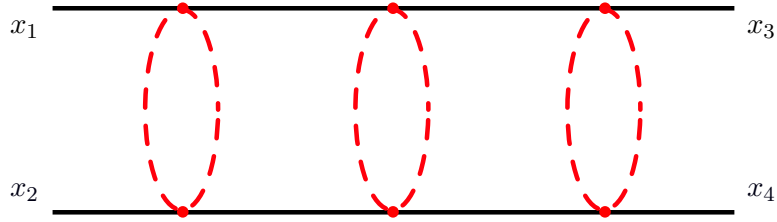
		\caption{An example of zero-magnon $u$-channel fishnet ladder diagram evaluated in the computation of $\cG^u(z,\zb)$. The dashed/solid four-point interaction sites have coupling $\xi^2$ so this diagram contributes at order $\xi^{12}$. The only other diagrams are of this form but with any number of ``rungs''.}
		\label{fishnet}
	}
\end{figure}

An example of a class of diagrams which have been resummed to all orders in the planar limit
are the ``fishnet'' diagrams drawn in figure \ref{fishnet}.
They describe the ``zero-magnon'' correlator (trace implied):
\be \label{zero magnon}
 \<0 | X^\dagger(x_4)X^\dagger(x_3) X(x_2)X(x_1)|0\> \equiv \frac{1}{x_{12}^2x_{34}^2}  \cG(z,\zb)
\ee
which was computed exactly in the $u$-channel in \cite{Gromov:2018hut} (eqs.~(3.12) and (A.6) there)
as a sum over conformal blocks
\begin{equation}\label{corr}
\cG^u(z,\zb)=\sum_{J\geq 0} (-1)^J \int_{-\infty}^{\infty} \frac{d\nu}{2\pi} C_{\Delta, J}\frac{256E_{\Delta,J}}{1-256\: \xi^4 E_{\Delta,J} } \: G^{(0,0)}_{\Delta,J}(z,\zb)\ ,
\end{equation}
where the scaling dimension is $\Delta = 2 + i\nu$,
\be\label{energyeqn}
E_{\Delta,J}=\frac{1}{16 (-\Delta+J+2)(-\Delta+J+4)(\Delta+J-2)(\Delta+J)}
\ee
and the normalization coefficient is \cite{Gromov:2018hut}%
\footnote{
We removed an overall factor of $1/(4\pi)^2$ in eq.~\eqref{zero magnon} and absorbed factors of $2$ and $\pi$ into $C$.}
%In addition, $C$ differs by a factor of $2$ from the definition in eq. (A.6) of \cite{Gromov:2018hut}, which we find is needed to match their other results. SCH: $d\nu^{\rm here}=2d\nu^{\rm there}$?}
\be
C_{\Delta, J} =  \frac{\Gamma(\Delta-1)\Gamma(2+J)\Gamma(\frac{1}{2}(\Delta+J))^2 \Gamma(4-\Delta+J)}
{2 \Gamma(J+1)\Gamma(\Delta-2) \Gamma(\Delta+J-1)\Gamma(2-\frac{1}{2}(\Delta-J))^2}\ .
\ee
The conformal block $G^{(a,b)}_{\Delta,J}(z,\zb)$ is a combination of hypergeometric functions in $d=4$, see appendix~\ref{app:blocks}.
For this calculation, we have $\Delta_1=\Delta_2=\Delta_3=\Delta_4=1$ and hence $a=b=0$.
The $t$-channel ladders are given by the same expression \eqref{corr} without the $(-1)^J$.
We also require $\xi^2$ to have a small and negative imaginary component for the $\nu$ integration to well-defined \cite{Gromov:2018hut}.

Eq.~\eqref{corr} can be related to the usual operator product expansion by noticing that since the conformal blocks decay exponentially as $\Ima(\nu)\to-\infty$, we can close the integration contour in the lower half-plane and apply Cauchy's residue theorem. The poles of the integrand occur when $\nu$ solves
\begin{equation} \label{const}
1-256\: \xi^4 E_{2+i\nu,J}=0,
\end{equation}
while all other poles are spurious and cancel in pairs \cite{Gromov:2018hut}.
The spectrum of this correlator thus consists of exactly two Regge trajectories: only two operators contribute for each spin.
This result is valid for any finite coupling $\xi$, and in particular to all orders in perturbation theory, where the correlator is expanded as
\begin{equation}\label{expansion}
\cG^u(z,\zb)=\frac{z \zb}{z-\zb}\sum_{n=0}^{\infty} (i \xi^2)^n \cG^{(n)}(z,\zb).
\end{equation}
By analyzing the series expansions in small $z$ and $\zb$,
the authors of \cite{Gromov:2018hut} found that $\cG^{(n)}$'s are combinations of single-valued harmonic polylogarithms (HPLs), a basis for special iterated integrals.
Several useful properties and definitions of these functions are reviewed in appendix \ref{app:HPL}.
For example,
\begin{equation}
\cG^{(0)}(z,\zb) = z-\zb
\end{equation} and the order $\xi^2$ contribution is
\begin{multline}
\cG^{(1)}(z,\zb) = \mathcal{L}_{1,0}(z,\zb)-\mathcal{L}_{1,0}(z,\zb) \equiv H_{1,0}(z)-H_{1,0}(\zb)+H_1(z)H_0(\zb)\\- H_0(z)H_1(\zb)+H_{0,1}(\zb) -H_{0,1}(z).
\end{multline}
This particular function can be written explicitly in terms of ordinary dilogarithms (see eq.~(\ref{logextraction})):\begin{equation}
\cG^{(1)}(z,\zb) = -2\Li(z)+2\Li(\zb)+\big(\log(1-\zb)-\log(1-z)\big)\log(z\zb),
\end{equation} where $\Li$ is the dilogarithm function.
We verified the formulas provided in ref.~\cite{Gromov:2018hut} up to 6-loops (order $\xi^{12}$) and order $\sigma^4$.

\begin{figure}[t]
	\def\svgwidth{\linewidth}
	\centering{
		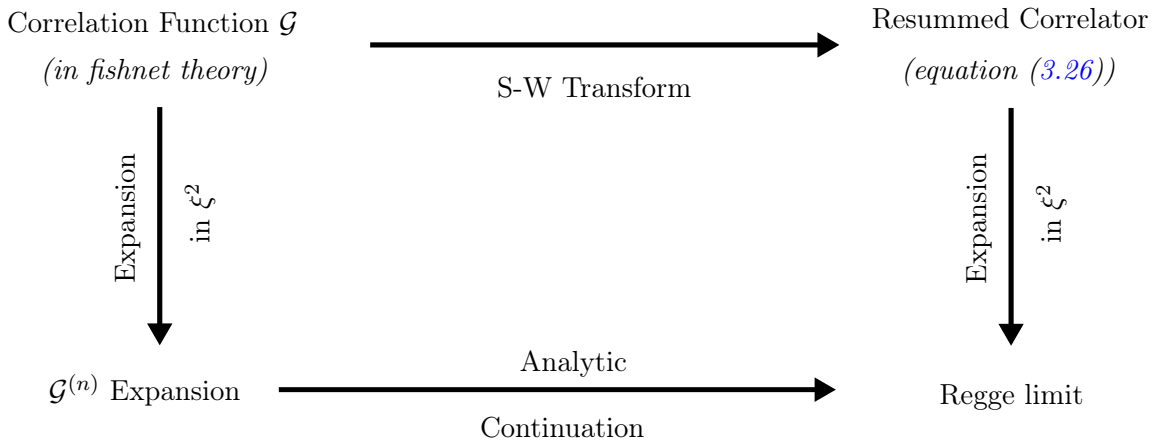
		\caption{A sketch of the processes evaluated in the following chapters.  We analyze the Sommerfeld-Watson resummation in generic conformal theories and demonstrate that the diagram commutes in the fishnet model.}
		\label{fig:topView}
	}
\end{figure}

The functions $\cG^{(L)}$ provide a ``data mine'' on which we can precisely test conformal Regge theory.
Regge theory allows us to resum the OPE beyond its radius of convergence in cross-ratio space
and to evaluate the correlator in the Regge limit via eq.~(\ref{Regge disc intro}).
Our first goal will be to check that this agrees, order by order in the coupling \emph{and} power by power in $\sigma$,
with the analytic continuation of the $\cG^{(L)}$'s.
A schematic of the calculation is provided in figure \ref{fig:topView}.
We also applied this technique to the ``one-magnon'' four-point function,
which has very similar structure to the zero-magnon case and is reviewed in section \ref{onemagnon}.
The Regge limit at leading power has been preceedingly studied in ref.~\cite{Korchemsky:2018hnb} and
was extended to other fishnet correlators in \cite{Chowdhury:2019hns,Chowdhury:2020tbn}.

\section{Conformal Regge theory with exact energy dependence \label{confregge}}

The extension of the $s$-channel OPE to the Regge limit described in section \ref{reggeReview}
was obtained in the seminal paper \cite{Costa:2012cb}.
This is nontrivial since the sum over spins diverges in the Regge regime. The solution
is to rewrite the sum as an integral via the so-called Sommerfeld-Watson transform.
Our contribution here will be to extend the formulas of \cite{Costa:2012cb} to an exact expression (see eqs.~\eqref{Regge disc}-\eqref{main} below)
which can be used to obtain arbitrary subleading powers of $z,\zb$.
As we will see, a new sort of term then appears.
In this section we keep the spacetime dimension and external operator dimensions generic.

\subsection{Sommerfeld-Watson resummation in S-matrix theory \label{confreggesmatrix}}

We begin by reviewing the classic resummation of SO($d$) spherical harmonics,
which will give us intuition about what we should, and \emph{should not}, expect
(see also \cite{Collins:1977jy,Donnachie:2002en}).
Consider a function of one angle:
\be
F(\cos\theta) = \sum_J a_J \P_J(\cos\theta) \label{flat OPE}
\ee
where the SO($d$) spherical harmonics $\P_J$ are defined in eq.~\eqref{def CJ}.
We use a normalization which trivializes the Regge limit: $\lim_{x\to\infty}\P_J(x) \to (2x)^J$.
We will borrow nomenclature from  S-matrix theory, where in $d$ spacetime dimensions
one would use SO($d{-}1$) partial waves, $\cos\theta=1+\tfrac{2t}{s}$ (say for massless scattering),
and the coefficients would depend on center-of-mass energy $s$.
Regge theory aims to use such $s$-channel partial waves
to study the large-$t$, fixed-$s$ Regge limit; the $s$-dependence will play no role in our discussion.

As reviewed below, in S-matrix applications
the partial waves $a_J$ are the sum of a part which is analytic and one which alternates with spin:
\be
 a_J = a_J^t + (-1)^J a_J^u\,, \label{atu}
\ee
where each of $a_J^{t,u}$ is analytic and polynomially bounded in a half-plane
${\rm Re}(J)>j_*$.
These are associated with $t$- and $u$-channel cuts, representing
singularities at positive and negative $x$, respectively.
Many references use instead even- and odd- combinations ($a_J^t\pm a_J^u$).

\begin{figure}[t]
	\def\svgwidth{1\linewidth}
	\centering{
		%% Creator: Inkscape 1.0 (4035a4f, 2020-05-01), www.inkscape.org
%% Accompanies image file 'SommerfeldTransform1.pdf' (pdf, eps, ps)
\begingroup%
  \makeatletter%
  \providecommand\color[2][]{%
    \errmessage{(Inkscape) Color is used for the text in Inkscape, but the package 'color.sty' is not loaded}%
    \renewcommand\color[2][]{}%
  }%
  \providecommand\transparent[1]{%
    \errmessage{(Inkscape) Transparency is used (non-zero) for the text in Inkscape, but the package 'transparent.sty' is not loaded}%
    \renewcommand\transparent[1]{}%
  }%
  \providecommand\rotatebox[2]{#2}%
  \newcommand*\fsize{\dimexpr\f@size pt\relax}%
  \newcommand*\lineheight[1]{\fontsize{\fsize}{#1\fsize}\selectfont}%
  \ifx\svgwidth\undefined%
    \setlength{\unitlength}{708.66141732bp}%
    \ifx\svgscale\undefined%
      \relax%
    \else%
      \setlength{\unitlength}{\unitlength * \real{\svgscale}}%
    \fi%
  \else%
    \setlength{\unitlength}{\svgwidth}%
  \fi%
  \global\let\svgwidth\undefined%
  \global\let\svgscale\undefined%
  \makeatother%
  \begin{picture}(1,0.28)%
    \lineheight{1}%
    \setlength\tabcolsep{0pt}%
    \put(0,0){\includegraphics[width=\unitlength,page=1]{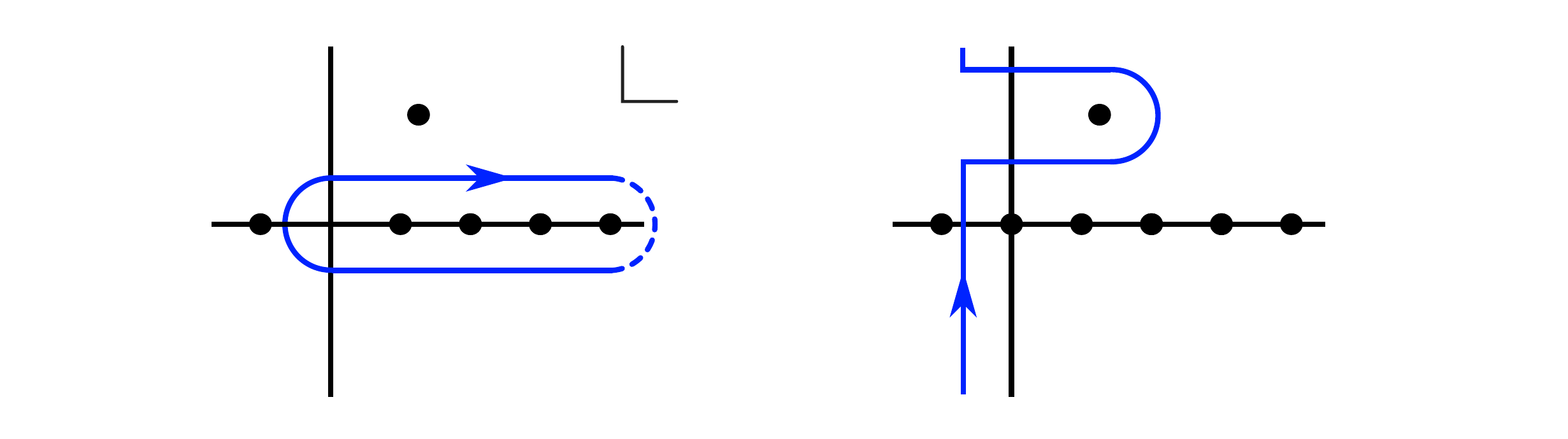}}%
    \put(0.41045044,0.2273078){\makebox(0,0)[lt]{\lineheight{1.25}\smash{\begin{tabular}[t]{l}$j$\end{tabular}}}}%
    \put(0.28307501,0.21112549){\makebox(0,0)[lt]{\lineheight{1.25}\smash{\begin{tabular}[t]{l}$j(\nu)$\end{tabular}}}}%
    \put(0.22023923,0.11211975){\makebox(0,0)[lt]{\lineheight{1.25}\smash{\begin{tabular}[t]{l}0\end{tabular}}}}%
    \put(0,0){\includegraphics[width=\unitlength,page=2]{SommerfeldTransform1.pdf}}%
    \put(0.26486722,0.11211975){\makebox(0,0)[lt]{\lineheight{1.25}\smash{\begin{tabular}[t]{l}1\end{tabular}}}}%
    \put(0,0){\includegraphics[width=\unitlength,page=3]{SommerfeldTransform1.pdf}}%
    \put(0.84466198,0.2273078){\makebox(0,0)[lt]{\lineheight{1.25}\smash{\begin{tabular}[t]{l}$j$\end{tabular}}}}%
    \put(0.37588702,0.08395422){\makebox(0,0)[lt]{\lineheight{1.25}\smash{\begin{tabular}[t]{l}$C$\end{tabular}}}}%
  \end{picture}%
\endgroup%

		\caption{First step of the Sommerfeld-Watson transform in S-matrix theory.
		The contour should remain to the right of $j$-plane singularities.
		The contours are equivalent in Euclidean kinematics but the second one allows a safe continuation to Lorentzian.}
		\label{fig:sommerfeld0}
	}
\end{figure}

To rewrite the sum \eqref{flat OPE} as an integral, we need to think of the analytic properties of
the spherical harmonics $\P_J(x)$.
These are entire functions of $J$ (except for the gamma function poles at negative $J$)
% at $J=2-d-m$ , $m=1,2,3\ldots$,
which generally have a ``$u$-channel'' cut for $x\in (\infty,-1]$.
In fact we have two natural functions: $\P_J(\pm x)$. They are related
by an overall sign $(-1)^J$ when the spin is an integer but generally they are distinct.
The Sommerfeld-Watson transform pairs $a_J^t$ with the function with a $t$-cut,
and $a_J^u$ with the function with a $u$-cut:
\be
 F(x) = -\pi\int_C \frac{dJ}{2\pi i}  \frac{a_J^t \P_J(-x) +a_J^u \P_J(x)}{\sin(\pi J)} + \mbox{(subtractions)} \label{flat WS}
\ee
where the contour $C$ encircles clockwise the poles of $1/\sin(\pi J)$ with $J\geq0$,
see fig.~\ref{fig:sommerfeld0}.
Since the residue of $1/\sin(\pi J)$ is proportional to $(-1)^J$, it is easy to verify
that the integral reproduces the sum in eq.~\eqref{flat OPE}.
The subtractions are a polynomial in $x$, accounting for the possibility
that for a finite number of spins the analytic continuation of $a_J^t$ may not agree with the
coefficient in eq.~\eqref{flat OPE}.

On the contour $C$, the integral (\ref{flat WS}) converges when the original sum does, {\it ie.} when $|\cos\theta|$ is not too large.
To gain anything from this trick one must deform the contour to a vertical line.
It will be convenient to center it on the fixed line of the Weyl reflection $J\mapsto 2-d-J$:
\be
 F(x) = -\pi \!\!\!\!\!\!\int\limits_{-\frac{d-2}{2}-i\infty}^{-\frac{d-2}{2}+i\infty}
 \frac{dJ}{2\pi i}  \frac{a_J^t \P_J(-x) +a_J^u \P_J(x)}{\sin(\pi J)}+ \mbox{(subtractions)} \,. \label{flat WS 1}
\ee
The contour should remain to the right of all singularities of the $a^{t,u}_J$.
In Euclidean kinematics, $\P_J(\cos\theta) \sim e^{\pm i\theta J}$ at large imaginary $J$
and the integral converges (possibly as a distribution)
as long as $\theta\in [0,\pi]$.  In Lorentzian kinematics with $|x|\gg 1$,
$\P_J(x)\sim (2x)^J$ and we retain convergence on this contour as long as $\arg(x)\in [0,\pi]$.
%Singularities near the origin will be discussed shortly.
We stress that, given $a_J^t$ and $a_J^u$, eq.~\eqref{flat WS 1} is an exact representation for the function $F(x)$.

Let us comment on the meaning of the coefficients $a_J^{t,u}$ in eq.~\eqref{flat WS 1}.
In general, they are analytic functions in some half-plane ${\rm Re}(J)\geq j_*$ which may not include
the vertical line ${\rm Re}(J)=-\frac{d-2}{2}$. In drawing the second contour in fig.~\ref{fig:sommerfeld0}
we assumed that singularities occur at finite ${\rm Im}\ J$, so that there are no obstructions to reaching the vertical line
at large imaginary $J$.
This seems physically reasonable since large-spin is often a semi-classical limit.
The same comment will apply below in CFT.

A typical application of eq.~\eqref{flat WS 1} is to obtain large-$x$ asymptotics.
The intermediate steps are subtle if one is interested in subleading terms,
but since the answer is surprisingly simple, it will be worth going through the steps (following appendix A of \cite{Donnachie:2002en}).
The basic idea is to split $\P_J(x)$
into two parts, which decay in the left and right $J$ half-planes respectively:
\be
 \P_J(x) = \Ppure_J(x) +\frac{\Gamma(J+d-2)\Gamma\big(-\tfrac{d-2}{2}-J\big)}{
 \Gamma\big(J+\tfrac{d-2}{2}\big)\Gamma(-J)} \Ppure_{2-d-j}(x)\,, \label{P from Ppures}
\ee
where $\Ppure_J(x)$ satisfies the same Casimir equation but contains a single tower of term as $x\to\infty$
(this function is proportional to Legendre's $Q$ when $d=3$):
\be
\Ppure_J(x) \equiv (2x)^J \ {}_2F_1\big(-\tfrac{J}{2},\tfrac{1-J}{2},2-\tfrac{d}{2}-J,\tfrac{1}{x^2}\big)\,.
\ee
We should thus deform the integration contour left for the $\Ppure_J$ terms, and right for $\Ppure_{2-d-J}$.
In principle, one has to include the following types of singularities:
\begin{enumerate}
\item Physical left poles or cuts from $a_j^{t,u}$.
\item Spurious left poles at $J=-\tfrac{d-2}{2}-m$, $m=1,2,3\ldots$ from $a_J^{t,u}$ (see below).
\item Right poles at $J=-\tfrac{d-2}{2}+m$, $m=1,2,3\ldots$ from gammas in eq.~\eqref{P from Ppures}.
\item Left poles at $J=-1,-2\ldots$ from $1/\sin(\pi J)$.
\item Right poles at $J=0,1,2\ldots$ from $1/\sin(\pi J)$.
\end{enumerate}
The surprise, remarkably, is that poles 2-5 all cancel out, and only the physical singularities of 
$a_j^{t,u}$ contribute!
In brief, poles $2$-$3$ are related by a Weyl reflection and cancel in pairs;
poles 4 are generically absent due to
a cancellation between $t$- and $u$-channel coefficients\footnote{
By ``generically'' we mean that
the combination $\frac{e^{-i\pi J} a_J^t +a_J^u}{\sin(\pi J)}$ has a pole at $J=-m$ with $m$=integer
if and only if the theory has a physical Regge pole at that location, corresponding to an integer-power term in the Regge limit,
which is not the case in a generic theory.},
and poles 5 are absent due to $1/\Gamma(-J)$ in eq.~\eqref{P from Ppures}.
The cancellations are detailed in appendix \ref{app:FG} and can be readily understood using a
concrete formula for the $a_J^{t,u}$, known as the Froissart-Gribov formula.

\begin{figure}[t]
	\def\svgwidth{0.6\linewidth}
	\centering{
		%% Creator: Inkscape 1.0 (4035a4f, 2020-05-01), www.inkscape.org
%% Accompanies image file 'SpuriousPoles2.pdf' (pdf, eps, ps)
\begingroup%
  \makeatletter%
  \providecommand\color[2][]{%
    \errmessage{(Inkscape) Color is used for the text in Inkscape, but the package 'color.sty' is not loaded}%
    \renewcommand\color[2][]{}%
  }%
  \providecommand\transparent[1]{%
    \errmessage{(Inkscape) Transparency is used (non-zero) for the text in Inkscape, but the package 'transparent.sty' is not loaded}%
    \renewcommand\transparent[1]{}%
  }%
  \providecommand\rotatebox[2]{#2}%
  \newcommand*\fsize{\dimexpr\f@size pt\relax}%
  \newcommand*\lineheight[1]{\fontsize{\fsize}{#1\fsize}\selectfont}%
  \ifx\svgwidth\undefined%
    \setlength{\unitlength}{311.81102362bp}%
    \ifx\svgscale\undefined%
      \relax%
    \else%
      \setlength{\unitlength}{\unitlength * \real{\svgscale}}%
    \fi%
  \else%
    \setlength{\unitlength}{\svgwidth}%
  \fi%
  \global\let\svgwidth\undefined%
  \global\let\svgscale\undefined%
  \makeatother%
  \begin{picture}(1,0.54545455)%
    \lineheight{1}%
    \setlength\tabcolsep{0pt}%
    \put(0,0){\includegraphics[width=\unitlength,page=1]{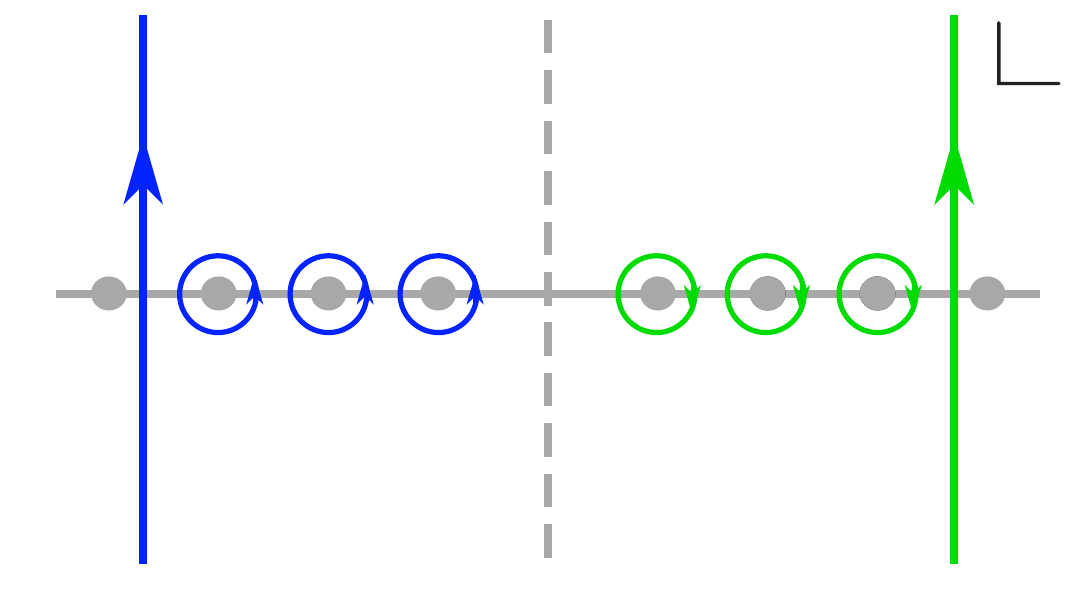}}%
    \put(0.94394934,0.48620494){\makebox(0,0)[lt]{\lineheight{1.25}\smash{\begin{tabular}[t]{l}$j$\end{tabular}}}}%
    \put(0,0){\includegraphics[width=\unitlength,page=2]{SpuriousPoles2.pdf}}%
    \put(0.38199466,0.05151173){\makebox(0,0)[lt]{\lineheight{1.25}\smash{\begin{tabular}[t]{l}$-\frac{d-2}{2}$\end{tabular}}}}%
  \end{picture}%
\endgroup%

		\caption{Spurious poles of type 2 and 3 cancel each other. The physical Regge pole is shaded black.}
		\label{fig:spurious1}
	}
\end{figure}

The upshot is that only physical singularities (as defined precisely in the appendix) contribute.
Considering, for notational simplicity, 
the case in which these consist of discrete poles at $J=j_n$,
and taking $x$ to be positive and above the real axis cut, the result is:
\be
\lim_{x\to\infty} F(x+i0) = -\pi \sum_{j_n} \Res\limits_{J=j_n}
\frac{e^{-i\pi J} a_J^t +a_J^u}{\sin(\pi J)}\Ppure_J(x)\,. \label{Regge}
\ee
This is a fundamental result of Regge theory.
Since $\Ppure_J(x)\approx (2x)^J\propto t^J$, the rightmost $J$-plane singularities dominate at $t\gg s$.
More generally, the sum gives an asymptotic expansion in $1/t$.
The phases of the two terms
in eq.~\eqref{Regge} are simply those of $({-}t{-}i0)^J$ and $({-}u{-}i0)^J$, respectively.

The remarkable feature of eq.~\eqref{Regge} is that, to correctly reproduce the amplitude
to any desired order in the $1/t$ expansion of $F$, it suffices to replace
$\P_J$ by $\Ppure_J$ in the Sommerfeld-Watson formula \eqref{flat WS 1} \emph{and ignore
spurious singularities of $1/\sin(\pi J)$ and $a_J^{t,u}$}.

\subsection{Analytic continuation to the Lorentzian regime and Regge block}

The spectral representation of correlation functions is the starting point for Regge analysis in conformal theories.
It allows to write the correlation function as an integral over continuous dimensions:
\begin{equation} \label{spectral}
\cG(z, \zb) = \sum_J \int_{-\infty}^{\infty} \frac{d\nu}{2\pi}\: c(\Delta,J) F^{(a,b)}_{\Delta,J}(z, \zb) +\mbox{(non-norm)},
\end{equation}
where the exchanged operator scaling dimension is parametrized as $\Delta=d/2+i\nu$, where $\nu$ is a complex number.
The meromorphic function $c(J, \Delta)$ contains the OPE coefficient data of a particular theory, and has poles at the location of local operators.
The ``non-normalizable'' modes account for operators with $\Delta<d/2$ (which includes, notably, the identity)
\cite{Simmons-Duffin:2017nub}.
The conformal partial waves $F_{J,\Delta}$ are a sum of conformal block and its shadow \cite{SimmonsDuffin:2012uy,Costa:2012cb,Caron-Huot:2017vep},
\begin{equation}\label{feqn}
F^{(a,b)}_{\Delta,J}(z,\zb) = \frac{1}{2}\bigg(G^{(a,b)}_{\Delta,J}(z,\zb) + \frac{K^{(a,b)}_{d-\Delta,J}}{K^{(a,b)}_{\Delta,J}}G^{(a,b)}_{d-\Delta,J}(z,\zb) \bigg),
\end{equation}
with coefficient that are products of gamma functions
\begin{equation}\label{kappaeqns}
K^{(a,b)}_{\Delta,J} = \frac{\Gamma(\Delta - 1)}{\Gamma(\Delta - \frac{d}{2})} \kappa^{(a,b)}_{\Delta+J}, \qquad \kappa^{(a,b)}_{\beta} = \frac{\Gamma(\frac{\beta}{2}-a)\Gamma(\frac{\beta}{2}+a)\Gamma(\frac{\beta}{2}-b)\Gamma(\frac{\beta}{2}+b)}{2 \pi^2 \Gamma(\beta-1)\Gamma(\beta)}\ .
\end{equation}

The spectral representation \eqref{spectral} involves a discrete sum over spins, analogous to eq.~\eqref{flat OPE}.
To reach the Lorentzian regime, we must first replace the sum by an integral, and then analytically continue
$\zb$ counterclockwise around 1.

This process has been discussed many times, but we found an unexpected twist:
the first step enjoys some freedom because one can add to $F_{\Delta,J}$
terms which vanish for integer spin. We find that the next steps is greatly simplified, especially at subleading powers,
by making such an improvement. This discussion will be somewhat technical.

Let us recall the defining properties of $F_{\Delta,J}$: it satisfies the same Casimir equation as $G_{\Delta,J}$,
and it is Euclidean single-valued (meaning, it has no branch cut when $\zb=z^*$).
The problem with eq.~\eqref{feqn} is that this property does not hold for non-integer spin---this combination is then
not natural in any sense!  In fact no combination of $G$'s can satisfy Euclidean single-valuedness for non-integer $J$,
because it is violated in the $z,\zb\to 0$ limit:
\be
\lim_{z\ll\zb\ll 1} G^{(a,b)}_{\Delta,J}= (z\zb)^{\Delta/2} (z/\zb)^{J/2}.  \label{limit G}
\ee
The last factor is only Euclidean single-valued when $J$ is an integer.
Our proposed resolution is that one can \emph{still} impose Euclidean single-valuedness in either the left or right half-plane.
Given our kinematics of interest, we pick the second option,
meaning in particular that we cancel the monodromy around the point $(z,\zb)=(1,1)$.
To construct the corresponding block, we make an ansatz using
the three functions: $G_{\Delta,J}$, $G_{d-\Delta,J}$ and
$G_{J+d-1,1 - \Delta}$. These solve the same Casimir equation, are regular at $z=\zb$,
and do not have contain singular powers $(z\zb)^{-J/2}$ at positive $J$.
Using the method detailed shortly, we find that the \emph{natural} non-integer spin version of eq.~\eqref{feqn} contains a third term:
\ba
 F^{(a,b)\rm good}_{\Delta,J}(z,\zb) &= 
 \frac{1}{2}G^{(a,b)}_{\Delta,J}(z,\zb) + \frac12\frac{K_{d-\Delta,J}}{K_{\Delta,J}}G^{(a,b)}_{d-\Delta,J}(z,\zb) +
\\ &\phantom{=} + \pi K_{J+d-1,1-\Delta} \frac{\Gamma\big(-J-\tfrac{d-2}{2}\big)}{\Gamma(-J)}\left( s^{(a,b)}_{\Delta+J}-s^{(a,b)}_{\Delta+2-d-J}\right) G^{(a,b)}_{J+d-1,1 - \Delta}(z,\zb)\,, \label{Fgood}
\ea 
where $s$ is a product of sines which will often reoccur:
\be
s^{(a,b)}_\beta = \frac{\sin\pi\big(\tfrac{\beta}{2}+a\big)\sin\pi\big(\tfrac{\beta}{2}+b\big)}{\sin(\pi\beta)}\,.
\ee
For the moment, we remark only that the second ine of eq.~\eqref{Fgood}
manifestly vanishes for integer $J\geq 0$, due to $1/\Gamma(-J)$, so $F^{\rm good}$ reduces to $F$ in that case.  Also, trigonometric identities can be used to show that the definition is invariant under the symmetry $(a,b)\mapsto(-a,-b)$.
To our knowledge, the function in eq.~\eqref{Fgood} is new. It would be interesting to interpret it in the language of
shadow representation, light transforms or integrability \cite{SimmonsDuffin:2012uy,Kravchuk:2018htv,Isachenkov:2017qgn},
and also to compare with the function called $\mathcal{G}$ in ref.~\cite{Raben:2018sjl}.

Our method to analytically continue $F^{(a,b)\rm good}_{\Delta,J}(z,\zb)$
to the Regge sheet, following the path in fig.~\ref{continuation},
is the same method that we used to find the coefficients in eq.~\eqref{Fgood}.
We first decompose each block $G^{(a,b)}_{\Delta,J}$ into pure power solutions according to
\begin{equation}\label{Gdecomp}
G^{(a,b)}_{\Delta,J}(z,\zb)=g_{\Delta,J}^{(a,b)\pure}(z,\zb)+\frac{\Gamma(J+d-2)\Gamma(-J-\frac{d-2}{2})}{\Gamma(J+\frac{d-2}{2})\Gamma(-J)} g_{\Delta,-J-d+2}^{(a,b)\pure}(z,\zb)\,,
\end{equation}
where each $g^{\pure}$ contains a single tower of terms in the limit (\ref{limit G}). This decomposition is identical to that
used for spherical harmonics in eq.~\eqref{P from Ppures}.
Contrary to $G$, the $g^\pure$'s are not symmetrical in $(z,\zb)$. They are however
easy to analytically continue around $\zb=1$: since $z$ is held fixed during the continuation, the exponent of $z$ cannot change \cite{Caron-Huot:2017vep, Costa:2012cb, Costa:2017twz}:
\begin{equation} \label{contg}
g_{\Delta,J}^{(a,b)\pure}(z,\zb^{\circlearrowleft}) = \left(1-2ie^{-i\pi(a+b)} s^{(a,b)}_{\Delta+J}\right)\: g_{\Delta,J}^{(a,b)\pure}(z,\zb)-\frac{i}{\pi}\frac{e^{-i\pi(a+b)}}{\kappa^{(a,b)}_{\Delta+J}} g_{1-\Delta,1-J}^{(a,b)\pure}(z,\zb)\, .
\end{equation}
The continuation of $F^{(a,b)}_{\Delta,J}(z,\zb)$ or $F^{(a,b)\rm good}_{J,\Delta}(z,\zb)$ leaves us with eight $g^{\pure}$'s with various complicated coefficients. 

\begin{figure}[t]
	\def\svgwidth{0.7\linewidth}
	\centering{
		%% Creator: Inkscape 1.0 (4035a4f, 2020-05-01), www.inkscape.org
%% Accompanies image file 'IFF.pdf' (pdf, eps, ps)
\begingroup%
  \makeatletter%
  \providecommand\color[2][]{%
    \errmessage{(Inkscape) Color is used for the text in Inkscape, but the package 'color.sty' is not loaded}%
    \renewcommand\color[2][]{}%
  }%
  \providecommand\transparent[1]{%
    \errmessage{(Inkscape) Transparency is used (non-zero) for the text in Inkscape, but the package 'transparent.sty' is not loaded}%
    \renewcommand\transparent[1]{}%
  }%
  \providecommand\rotatebox[2]{#2}%
  \newcommand*\fsize{\dimexpr\f@size pt\relax}%
  \newcommand*\lineheight[1]{\fontsize{\fsize}{#1\fsize}\selectfont}%
  \ifx\svgwidth\undefined%
    \setlength{\unitlength}{510.23622047bp}%
    \ifx\svgscale\undefined%
      \relax%
    \else%
      \setlength{\unitlength}{\unitlength * \real{\svgscale}}%
    \fi%
  \else%
    \setlength{\unitlength}{\svgwidth}%
  \fi%
  \global\let\svgwidth\undefined%
  \global\let\svgscale\undefined%
  \makeatother%
  \begin{picture}(1,0.26666667)%
    \lineheight{1}%
    \setlength\tabcolsep{0pt}%
    \put(0,0){\includegraphics[width=\unitlength,page=1]{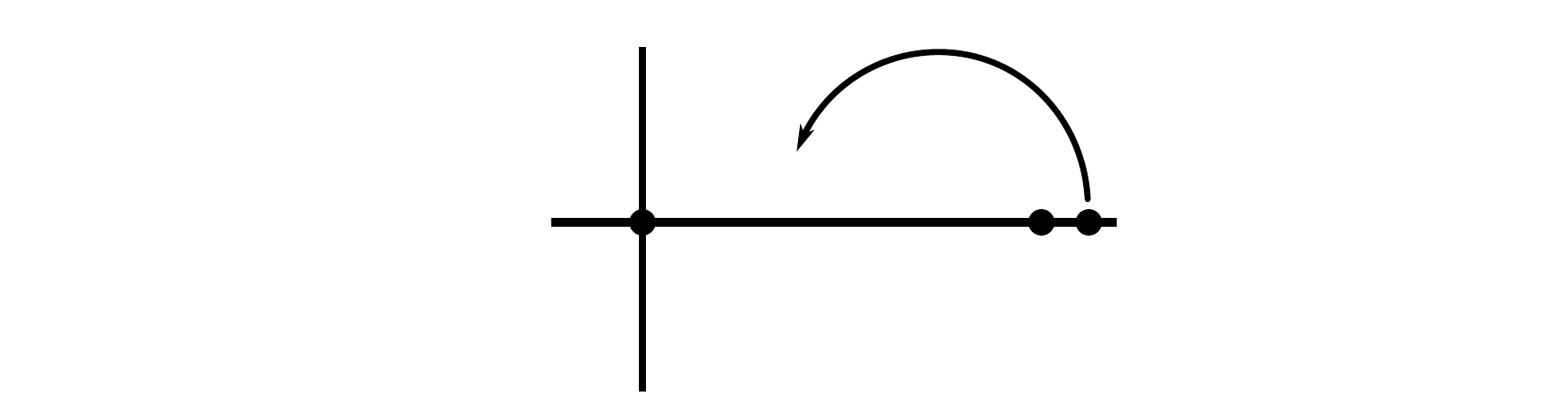}}%
    \put(0.4191939,0.08621063){\color[rgb]{0,0,0.34509804}\makebox(0,0)[lt]{\lineheight{1.25}\smash{\begin{tabular}[t]{l}0\end{tabular}}}}%
    \put(0.6077318,0.08327077){\color[rgb]{0,0,0.34509804}\makebox(0,0)[lt]{\lineheight{1.25}\smash{\begin{tabular}[t]{l}1\end{tabular}}}}%
    \put(0,0){\includegraphics[width=\unitlength,page=2]{IFF.pdf}}%
    \put(0.5998216,0.15882017){\color[rgb]{0,0.00392157,0.2627451}\makebox(0,0)[lt]{\lineheight{1.25}\smash{\begin{tabular}[t]{l}$z$\end{tabular}}}}%
    \put(0.52831438,0.23277902){\color[rgb]{0,0.00392157,0.2627451}\makebox(0,0)[lt]{\lineheight{1.25}\smash{\begin{tabular}[t]{l}$\bar{z}$\end{tabular}}}}%
    \put(0,0){\includegraphics[width=\unitlength,page=3]{IFF.pdf}}%
  \end{picture}%
\endgroup%

		\caption{Rotation of $z,\bar{z}$ counterclockwise from $z,\bar{z}>1$.
			\label{continuationIFF}
	}}
\end{figure}

Now the crux is that a combination of blocks $F(z,\zb)$ is Euclidean single-valued around $(1,1)$
\emph{if and only if} the $g^{\rm pure}$'s can be re-packaged into $G$'s. The reason is that we can reach the Regge sheet by rotating $z,\zb$ counter-clockwise starting from the region $z,\zb>1$, where Euclidean single-valued functions are symmetrical: $F(z+i0,\zb-i0)=F(z-i0,\zb+i0)$ (see fig. \ref{continuationIFF}).  Since we can reach the Regge sheet by continuing $z$, $\zb$ symmetrically from a region where the correlator is symmetrical, it follows that the continuation of a single-valued correlator
is also symmetrical: $F(z,\zb^\circlearrowleft)=F(z^\circlearrowleft,\zb)$ (and nonsingular at $z=\zb$).
This property ensures that it is a sum of $G$'s.
This property fails for non-integer spins for the combination $F^{(a,b)}_{\Delta,J}$,
but it is restored by the unique combination $F^{(a,b)\rm good}_{\Delta,J}$.
This is how we determined eq.~\eqref{Fgood}.

The coefficients of the four resulting $G$'s contain a part that is essentially  the original
$F^{(a,b)\rm good}_{\Delta,J}$.
We thus subtract those off and record the discontinuity:
\ba
\Disc_{14} F^{(a,b)\rm good}_{\Delta,J}(z,\zb) &= 
-i\left(e^{i\pi(a+b)}F^{(a,b)\rm good}_{\Delta,J}(z,\zb^\circlearrowleft) -e^{-i\pi(a+b)}F^{(a,b)\rm good}_{\Delta,J}(z,\zb)\right)
\\
&\equiv -\frac{R^{(a,b)}_{\Delta,J}(z,\zb)}{2\pi \kappa^{(a,b)}_{\Delta+J}} \,, \label{defR}
\ea
which is given in terms of a new ``Regge block'':
\ba
 R^{(a,b)}_{\Delta,J} &= G^{(a,b)}_{1-J,1-\Delta}
 -\kappat^{(a,b)}_{\Delta+J}\ G^{(a,b)}_{\Delta,J}
-\frac{\Gamma(d-\Delta-1)\Gamma\big(\Delta-\tfrac{d}{2}\big)}{\Gamma(\Delta-1)\Gamma\big(\tfrac{d}{2}-\Delta\big)}\kappat^{(a,b)}_{d-\Delta+J}\ G^{(a,b)}_{d-\Delta,J}+
\\ &\phantom{=}+ 
\frac{\Gamma(J+d-2)\Gamma\big(-J-\tfrac{d-2}{2}\big)}{\Gamma\big(J+\tfrac{d-2}{2}\big)\Gamma(-J)}
\kappat^{(a,b)}_{\Delta+J}\ \kappat^{(a,b)}_{d-\Delta+J}\ G^{(a,b)}_{J+d-1,1-\Delta}\,. \label{Rab}
\ea
Here we defined the following product of $\Gamma$-functions:
\be \label{kappat}
 \kappat^{(a,b)}_\beta = \frac{r^{(a,b)}_\beta}{r^{(a,b)}_{2-\beta}}\quad\mbox{with}\quad
 r^{(a,b)}_\beta \equiv \frac{\Gamma\big(\tfrac{\beta}{2}+a\big)\Gamma\big(\tfrac{\beta}{2}+b\big)}{\Gamma(\beta)}\ .
\ee
Importantly, the continuation (\ref{defR}) is exact even for noninteger $J$.

The first term of (\ref{Rab}) dominates in the Regge limit, $G^{(a,b)}_{1-J,1-\Delta}\sim \sigma^{1-J}$,
and it (deservedly) receives the most attention \cite{Costa:2012cb,Costa:2017twz,Kravchuk:2018htv}. 
However we will find that the other terms contribute nontrivially at subleading powers.

A simple defining property of $R^{(a,b)}_{\Delta,J}$ is that, being a discontinuity of blocks,
its other discontinuity vanishes:
\be
 \Disc_{23} R^{(a,b)}_{\Delta,J}(z,\zb) = 0. \label{disc R}
\ee
We find that eq.~\eqref{Rab} is the only combination of $G$'s satisfying this.
Alternatively, one could have defined a Regge block by taking the other discontinuity,
$R'\propto \Disc_{23}F^{(a,b)\rm good}_{\Delta,J}$:
this is given by the same expression \eqref{Rab} but with $\kappat^{(-a,-b)}_\beta$.

\subsection{Sommerfeld-Watson transformation}\label{sommerfeld}

With the analytic continuation of blocks worked out, one can try to evaluate
the continued correlation function following the path fig.~\ref{continuation} and discontinuity:
\begin{equation}
%\mathcal{A}(z, \zb) \equiv
\Disc_{14}\cG(z,\zb) \stackrel{?}{=}
\sum_{J\geq 0}^{\infty} \int_{-\infty}^{\infty} \frac{d\nu}{2\pi}\: c(\Delta,J) \frac{R^{(a,b)}_{\Delta,J}(z,\zb)}{2\pi \kappa^{(a,b)}_{\Delta+J}} \,.
\end{equation}
After some inspection, one finds that this expression makes no sense:
the Regge block $R_{\Delta,J}$ scales as $\sigma^{1-J}$ as $\sigma\to 0$ in the Regge limit, so the sum diverges.

Just as for the S-matrix Regge limit, the solution is to step back and rewrite the sum as an integral \emph{before} analytically continuing the cross-ratios to take the discontinuity.
This requires first promoting the spin $J$ to a complex variable and the partial wave coefficient $c(J,\Delta)$ to analytic functions of $J$.
In the S-matrix case, this possibility was first observed by Regge and was soon proved generally by Froissart and Gribov;
the analogous result in CFT was proved recently \cite{Caron-Huot:2017vep,Simmons-Duffin:2017nub,Kravchuk:2018htv}.
In general, this works for $J>j_*$ where it is known that $j_*\leq 1$ in a unitary theory.

The partial waves form not one, but in fact two analytic functions of spin:
\be\label{partialwavesplitting}
 c(\Delta,J) = c^t(\Delta,J) + (-1)^J c^u(\Delta,J),
\ee
where each term is nicely behaved (power-law bounded) at large imaginary $J$.

\begin{figure}[t]
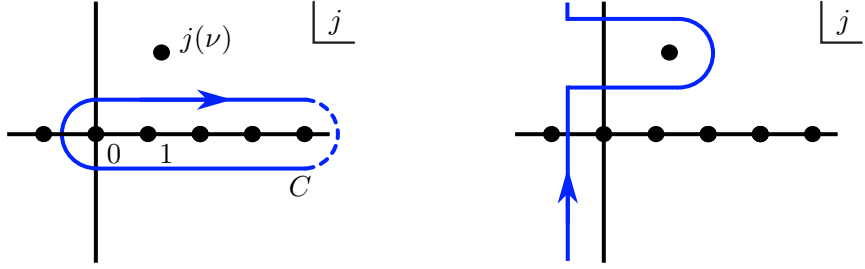

	\def\svgwidth{1\linewidth}
	\centering{
		
		\caption{Similar to fig.~\ref{fig:sommerfeld0}: The Sommerfeld-Watson transform for the
		$F^{(a,b)\rm good}_{\Delta,J}(z, \zb) $ block. 
		The contour should remain to the right of singularities of $c^{t,u}$.
		The contours are equivalent in Euclidean kinematics, and the second one allows a safe continuation to Lorentzian.}
		\label{fig:sommerfeld1}
	}
\end{figure}

Regge's idea allows us to express the sum over integer spins as an integral in the complex plane,
\begin{equation} \label{useless contour}
\cG(z, \zb) = -\pi \int_C \frac{dJ}{2\pi i} \int_{-\infty}^{\infty} \frac{d\nu}{2\pi}\: \frac{e^{i\pi J}c^t(\Delta,J) + c^u(\Delta,J)}{\sin(\pi J)} F^{(a,b)\rm good}_{\Delta,J}(z, \zb) %^{\circlearrowleft},
\end{equation}
where the contour $C$ envelopes the positive real $j$ axis, as illustrated in figure \ref{fig:sommerfeld1}.
Once this contour is in place, we can drag it around the complex plane to obtain a form where analytic continuation is possible.
The general technique is known as the Sommerfeld-Watson transform.

In the contour deformation of fig.~\ref{fig:sommerfeld1} we may encounter poles from the coefficients $c^{t,u}(\Delta,J)$,
as well as possible spurious poles from $F$.  Such spurious poles were discussed in \cite{Cornalba:2007fs}.
However, we find that these are absent when using the block $F^{\rm good}$, for a simple reason:
as the unique Casimir eigenfunction satisfying certain regularity conditions, $F^{\rm good}_{\Delta,J}$ is automatically
analytic for ${\rm Re}(J)> -\frac{d-2}{2}$ when $\Delta$ is along
the principal series ${\rm Re}(\Delta)=\tfrac{d}{2}$. We have also verified explicitly the
cancellation of poles using residue formulas from \cite{Kos:2013tga}.
We can thus write eq.~\eqref{useless contour} with a vertical contour:
\begin{equation}
\cG(z, \zb) = -\pi\!\!\!\!\!\int\limits_{-\frac{d-2}{2}-i\infty}^{-\frac{d-2}{2}+i\infty} \frac{dJ}{2\pi i}
\int\limits_{-\infty}^{\infty} \frac{d\nu}{2\pi}\: \frac{e^{i\pi J}c^t(\Delta,J) + c^u(\Delta,J)}{\sin(\pi J)} F^{(a,b)\rm good}_{\Delta,J}(z, \zb)\,. %^{\circlearrowleft},
\end{equation}
%(The contour should avoid from the right the singularities of  $c^{t,u}(\Delta,J)$.)
On this contour we are now allowed to analytically continue to the Regge sheet.
In particular we can take the discontinuity directly under the integration sign to get the Regge block in eq.~\eqref{Rab}:
\begin{equation}
\Disc_{14}\cG(z, \zb) = \int\limits_{-\frac{d-2}{2}-i\infty}^{-\frac{d-2}{2}+i\infty} \frac{dJ}{2\pi i}
\int\limits_{-\infty}^{\infty} \frac{d\nu}{2\pi}\: \frac{e^{i\pi J}c^t(\Delta,J) + c^u(\Delta,J)}{\sin(\pi J)}
\frac{R^{(a,b)}_{\Delta,J}(z,\zb)}{2\kappa^{(a,b)}_{\Delta+J}} +\mbox{(subtractions)}\,. %^{\circlearrowleft},
\label{Regge disc}
\end{equation}
We note that the sign of the phase $e^{i\pi J}$ is opposite in coordinate space as in momentum space.
The sign is forced on us since during the continuation,
the block contains a factor $F^{(a,b)\rm good}_{\Delta,J}(z, 1{-}i\epsilon)\sim e^{-\sqrt{i\epsilon} J}$
which grows at positive imaginary $J$. With the wrong choice $(-1)^J \mapsto e^{-i\pi J}$,
the integral would diverge. 

Eq.~(\ref{Regge disc}) is a central result of this paper: an \emph{exact} representation of the correlator on the Regge sheet.
This is a critical step toward obtaining asymptotic expansions, to which we now turn.
The ``subtractions'' include the discontinuity of non-normalizable modes in eq.~\eqref{spectral},
and possible low-spin corrections as in eq.~\eqref{flat WS}.

%We can now begin to evaluate this integral in the Regge limit: small $\sigma= \sqrt{z\zb}$ whilst maintaining fixed $w=\exp(\rho)= \sqrt{z/\zb}$.
%First, recall from eq.~\eqref{Rab} that $F_{J,\Delta}(z, \zb)^{\circlearrowleft}$ is a sum of four conformal blocks whose physics will be interpreted differently.  

\begin{figure}[t]
	\def\svgwidth{0.6\linewidth}
	\centering{
		%% Creator: Inkscape 1.0 (4035a4f, 2020-05-01), www.inkscape.org
%% Accompanies image file 'SpuriousPoles.pdf' (pdf, eps, ps)
\begingroup%
  \makeatletter%
  \providecommand\color[2][]{%
    \errmessage{(Inkscape) Color is used for the text in Inkscape, but the package 'color.sty' is not loaded}%
    \renewcommand\color[2][]{}%
  }%
  \providecommand\transparent[1]{%
    \errmessage{(Inkscape) Transparency is used (non-zero) for the text in Inkscape, but the package 'transparent.sty' is not loaded}%
    \renewcommand\transparent[1]{}%
  }%
  \providecommand\rotatebox[2]{#2}%
  \newcommand*\fsize{\dimexpr\f@size pt\relax}%
  \newcommand*\lineheight[1]{\fontsize{\fsize}{#1\fsize}\selectfont}%
  \ifx\svgwidth\undefined%
    \setlength{\unitlength}{311.81102362bp}%
    \ifx\svgscale\undefined%
      \relax%
    \else%
      \setlength{\unitlength}{\unitlength * \real{\svgscale}}%
    \fi%
  \else%
    \setlength{\unitlength}{\svgwidth}%
  \fi%
  \global\let\svgwidth\undefined%
  \global\let\svgscale\undefined%
  \makeatother%
  \begin{picture}(1,0.54545455)%
    \lineheight{1}%
    \setlength\tabcolsep{0pt}%
    \put(0,0){\includegraphics[width=\unitlength,page=1]{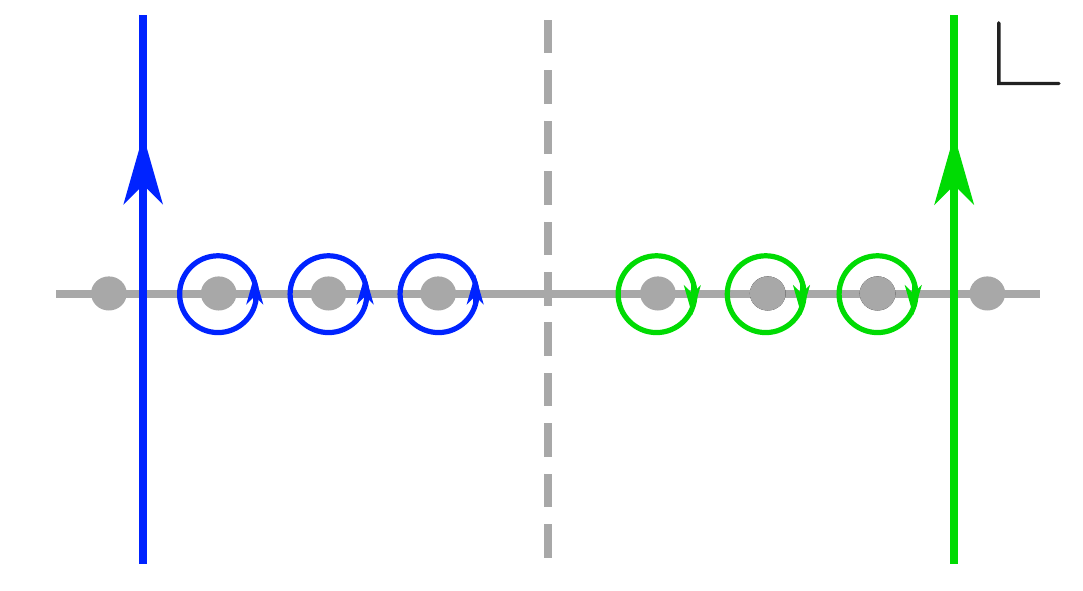}}%
    \put(0.94394934,0.48620494){\makebox(0,0)[lt]{\lineheight{1.25}\smash{\begin{tabular}[t]{l}$j$\end{tabular}}}}%
    \put(0,0){\includegraphics[width=\unitlength,page=2]{SpuriousPoles.pdf}}%
    \put(0.39161588,0.02264765){\makebox(0,0)[lt]{\lineheight{1.25}\smash{\begin{tabular}[t]{l}$-\frac{d-2}{2}$\end{tabular}}}}%
  \end{picture}%
\endgroup%

		\caption{Spurious poles of type 2-3 cancel against each other just as in the S-matrix case (fig. \ref{fig:spurious1}). Again, a physical Regge pole is shaded black.}
		\label{fig:spurious2}
	}
\end{figure}

Similar to eq.~\eqref{P from Ppures},
to derive an asymptotic expansion in the Regge limit $\sigma\to 0$
we simply write the Regge block as a term which decays on the left $J$-plane, plus
the three remaining $G$'s in eq.~\eqref{Rab}:
\be
 R^{(a,b)}_{\Delta,J}(z,\zb) = G^{(a,b)}_{1-J,1-\Delta}(z,\zb) + \mbox{rest}.
\ee
We then deform the $J$-contour left on the first term, and right for the rest.
Similar to section \ref{confreggesmatrix}, we find the following types of poles:
\begin{enumerate}
\item Physical left poles or cuts from $c^{t,u}(\Delta,J)$.
\item Spurious left poles at $J=-\tfrac{d-2}{2}-m$, $J=1-\Delta-m$ and $J=1-\tilde{\Delta}-m$,
$m=1,2,3\ldots$ from $C^{t,u}(\Delta,J)$.
\item Right poles at $J=-\tfrac{d-2}{2}+m$, $J=1-\Delta+m$ and $J=1-\tilde\Delta+m$ from rest of eq.~\eqref{Rab}
%at $J=-\tfrac{d-2}{2}+m$, $m=1,2,3\ldots$ from gammas in eq.~\eqref{P from Ppures}.
\item Left poles at $J=-1,-2\ldots$ from $1/\sin(\pi J)$.
\item Right poles at $J=0,1,2\ldots$ from $1/\sin(\pi J)$.
\end{enumerate}
Poles of types 2-4 cancel by the same two mechanisms discussed above. Namely, types
2-3 cancel among Weyl-reflected pairs $J\mapsto 2-J-d$ (see fig. \ref{fig:spurious2}),
by the mechanism detailed in eq.~\eqref{spurious 23}. The crux
is that argument is that the Regge block $R^{(a,b)}$ is free of spurious poles.
Type 4 poles multiply an explicit zero in the Lorentzian inversion formula, and so are generically
absent in the sense discussed in the S-matrix case.
Poles of type 5 however do \emph{not} cancel in the CFT case and must be retained.
The result is the following asymptotic expansion in the Regge limit, including subleading powers:
\ba \label{main}
\lim_{\sigma\to 0} \Disc_{14}\cG(z,\zb) &=
\int_{-\infty}^{\infty} \frac{d\nu}{2\pi} \sum_{j_n(\Delta)} \Res_{J=j_n(\Delta)}
\frac{e^{i\pi J}c^t(\Delta,J) + c^u(\Delta,J)}{\sin(\pi J)}
\frac{G^{(a,b)}_{1-J,1-\Delta}(z,\zb)}{2\kappa^{(a,b)}_{\Delta+J}}
\\ &\phantom{=} -
\sum\limits_{J\geq0} \int_{-\infty}^{\infty} \frac{d\nu}{2\pi} c(\Delta,J) s^{(-a,-b)}_{\Delta+J}2F_{\Delta,J}^{\prime(a,b)}(z,\zb)\,.
\ea
The first line could have been easily guessed and is as in S-matrix Regge theory (see eq.~\eqref{Regge}).
The second line is a new contribution which to our knowledge has not been discussed explicitly before;
it is important at subleading orders.  The block $F^\prime$ is defined similarly to
eq.~\eqref{feqn} with $\kappa\mapsto \kappat$ from eq.~\eqref{kappat}:
\be
 F_{\Delta,J}^{\prime(a,b)}(z,\zb) =\frac12\left( G_{\Delta,J}^{(a,b)}(z,\zb) +
 \frac{\kappat^{(a,b)}_{d-\Delta+J}}{\kappat^{(a,b)}_{\Delta+J}}
 \frac{\Gamma(d-\Delta-1)\Gamma\big(\Delta-\tfrac{d}{2}\big)}{\Gamma(\Delta-1)\Gamma\big(\tfrac{d}{2}-\Delta\big)}
 G_{d-\Delta,J}^{(a,b)}(z,\zb)\right).
\ee

Eqs.~\eqref{Regge disc} and \eqref{main} constitute the main results of this paper:  an exact expression
for correlators in Regge kinematics, and a corresponding all-order asymptotic expansion in the Regge limit.
The latter will be confronted in the next section with explicit expressions in the fishnet model.

\subsection{Formula for double-discontinuity: Recovering Lorentzian inversion}

As a first test of eq.~\eqref{Regge disc} we will now verify that it is consistent with the Lorentzian inversion formula.
The Lorentzian inversion formula extracts the OPE data from the double discontinuity:
\be
 c^t(\Delta,J) = \frac{\kappa^{(a,b)}_{\Delta+J}}{4} \int_0^1 dzd\zb \mu(z,\zb) G^{(-a,-b)}_{J+d-1,\Delta+1-d}(z,\zb) \dDisc[\cG(z,\zb)] \label{LIF}
\ee 
where the measure is $\mu=\frac{1}{z^2\zb^2} \left| \frac{z-\zb}{z\zb}\right|^{d-2}$ and the double-discontinuity is defined
as
\ba\label{dDisc def}
 \dDisc[\cG(z,\zb)] &\equiv \cos(\pi(a+b)) \cG(z,\zb) - \tfrac12e^{i\pi(a+b)}\cG(z,\zb^\circlearrowleft)
 - \tfrac12e^{-i\pi(a+b)}\cG(z,\zb^\circlearrowright)
 \\ &=\frac{-i}{2} \left( \Disc_{14}[\cG(z,\zb)] -\overline{\Disc}_{14}[\cG(z,\zb)]\right)
\ea
where the single discontinuity as defined in eq.~(\ref{Disc}) and $\overline{\Disc}$ is
the opposite analytic continuation with $i\mapsto -i$.
On the other hand, we just obtained an exact formula (\ref{Regge disc}) for the discontinuity of the correlator.
One might think that the double discontinuity should vanish since $\dDisc\propto \Disc_{23}\Disc_{14}$ which vanishes for any block, however, as stressed below eq.~\eqref{Regge disc},
the phase $e^{i\pi J}$ is only valid for the counter-clockwise path.
It is easy to see from the second form of $\dDisc$ that the dDisc is just the imaginary part of that phase, so that
the $c^u$ term and sine denominator simply cancel out:
\be
\label{Regge dDisc}
  \dDisc[\cG(z,\zb)] =
\int\limits_{-\frac{d-2}{2}-i\infty}^{-\frac{d-2}{2}+i\infty} \frac{dJ}{2\pi i}
\int\limits_{-\infty}^{\infty} \frac{d\nu}{2\pi}\: \frac{c^t(\Delta,J)}{2\kappa^{(a,b)}_{\Delta+J}}R^{(a,b)}_{\Delta,J}(z,\zb)\ .
\ee
This is the main result of this subsection. A similar formula was used recently in a paper involving one of the authors \cite{Caron-Huot:2020ouj}, but using only the $G^{(a,b)}_{1-J,1-\Delta}(z,\zb)$ part of the Regge block \eqref{Rab},
which was valid since that reference only considered the leading power. In contrast, eq.~\eqref{Regge dDisc} is an exact
representation.

Following the method above eq.~\eqref{main}, eq.~\eqref{Regge dDisc} can be used to obtain
asymptotic expansions in the Regge limit.  The difference between the formula with the $R$ block
and $G$ block is a function for which the $J$ contour can be deformed to the right, and whose purpose
is to cancel  type-2 spurious poles on the left. Thus eq.~\eqref{Regge dDisc} with
$R^{(a,b)}_{\Delta,J}(z,\zb)\mapsto G^{(a,b)}_{1-J,1-\Delta}(z,\zb)$ would only be valid if supplemented by an instruction
to discard type-2 spurious poles, and would then define only an asymptotic expansion.
The exact integral representation requires the Regge block $R$.

As a check, it is tempting to view eq.~\eqref{Regge dDisc} as the ``forward'' version of the Lorentzian inversion formula,
with the $G$ and $R$ blocks dual to each other. This requires the following pairing
to act as an orthogonality relation of sorts:
\be
P^{(a,b)}_{\Delta,J; \Delta',J'} \equiv
\frac{\kappa_{\Delta+J}^{(a,b)}}{4}
\int_0^1 dzd\zb\ \mu(z,\zb)\ G^{(-a,-b)}_{J+d-1,\Delta+1-d}(z,\zb) \frac{R^{(a,b)}_{\Delta,J}(z,\zb)}{2\kappa_{\Delta'+J'}^{(a,b)}}\,. \label{pairing}
\ee
In appendix \ref{app:Plancherel} we compute the integral exactly in $d=2$ and $d=4$, using the fact that
it factorizes into one-dimensional pairings which we could compute exactly using the Casimir equation satisfied by the blocks.
We find that in both dimensions the pairing is given by the following single formula:
\ba
P^{(a,b)}_{\Delta,J; \Delta',J'}&=
 \frac{r_{J+d-\Delta}^{(a,b)}}{r_{J'+d-\Delta'}^{(a,b)}}\frac{r_{J'+\Delta'}^{(-a,-b)}}{r_{J+\Delta}^{(-a,-b)}}
 \frac{4(J+\Delta-1) (J'+ d - \Delta' - 1)}{(\Delta{-}\Delta'{+}J{-}J')(\Delta{-}\Delta'{-}J{+}J')(\Delta{-}\tilde{\Delta}'{-}J{+}\tilde{J'})
(\Delta{-}\tilde{\Delta}'{+}J{-}\tilde{J'})}\\
&\phantom{=} + \mbox{($\Delta'$ shadow)}
\label{pairing answer text}
\ea
where $\tilde\Delta=d-\Delta$ and $\tilde J=2-d-J$ denote the dimension and spin shadows, respectively.
The last line is equal to the first with $\Delta'\mapsto \tilde\Delta'$
and multiplied by $K_{\tilde\Delta',J'}^{(a,b)}/K_{\Delta',J'}^{(a,b)}$, which is the appropriate relation
between the block and its shadow.

It would be interesting to compute eq.~\eqref{pairing answer text} in other spacetime dimensions.
Eq.~\eqref{pairing answer text} can't be quite the full answer when $d\neq 2,4$,
since in these cases it does not transform correctly under either $\Delta$ or $J'$ shadow transformations.

Plugging the Regge limit in eq.~\eqref{Regge dDisc} into the Lorentzian inversion formula
in eq.~\eqref{LIF}, the pairing should in principle recover the OPE data:
\be
 c^t(\Delta,J) \stackrel{?}{=} \int\limits_{-\frac{d-2}{2}-i\infty}^{-\frac{d-2}{2}+i\infty} \frac{dJ'}{2\pi i}
\int\limits_{\frac{d}{2}-i\infty}^{\frac{d}{2}+i\infty}
\frac{d\Delta'}{2\pi i} c^t(\Delta',J') P^{(a,b)}_{\Delta,J; \Delta',J'}\,.
\ee
Thanks to shadow symmetry of the coefficients $c^t(\Delta',J')$, we can ignore the second line
of eq.~\eqref{pairing answer text} at the cost of a factor 2.
In the Lorentzian inversion formula we are supposed to take $J>j_*$ so that the integral converges.
When ${\rm Re}(\Delta)=-d/2$, there are then no $J'$-poles in the right half-plane from
$c^t(\Delta',J')$ nor from $r$ functions, and the only singularities are two explicit poles
in $P^{(a,b)}_{\Delta,J; \Delta',J'}$. Deforming the $J'$ contour to the right we thus get:
\ba
c^t(\Delta,J) &\stackrel{?}{=}
\int\limits_{\frac{d}{2}-i\infty}^{\frac{d}{2}+i\infty} \frac{d\Delta'}{2\pi i(\Delta-\Delta')}
\Bigg[
\frac{r^{(-a,-b)}_{J-\Delta+2\Delta'}}{r^{(-a,-b)}_{J+\Delta}}\frac{J+\Delta-1}{J+\Delta'-1}c^t(\Delta',J-\Delta+\Delta')
\\&\hspace{35mm}
-\frac{r^{(a,b)}_{J+d-\Delta}}{r^{(a,b)}_{J+d+\Delta-2\Delta'}}\frac{J+d+\Delta-2\Delta'-1}{J+d-\Delta'-1}
c^t(\Delta',J+\Delta-\Delta')
\Bigg]\,.
\ea
We are reduced to a single integral.
Note that the two terms in the parenthesis cancel out when $\Delta'=\Delta$ so there are no singularities along
the integration contour.
To perform the integral, we notice that the top line is devoid of singularities
in the right half-plane ${\rm Re}\ \Delta'>d/2$, since the coefficient
$c^t(\Delta',J')$ is analytic between there and the unitarity bound, and the twist $\Delta'-J'=\Delta-J$ is held
constant (and below the unitarity bound for sufficiently large $J$) during integration.
Similarly, using the shadow relation between $c^t(\Delta',J'$) and $c^t(d-\Delta',J'$) the
second is devoid of poles in the left half-plane in $d=2,4$.  Starting from a contour slightly to the left, and deforming the contour in the two lines to the right and left, respectively, we thus pick a single pole from the top line:
\ba
c^t(\Delta,J) &\stackrel{?}{=} -\Res\limits_{\Delta'=\Delta} \frac{1}{\Delta-\Delta'}\
\frac{r^{(-a,-b)}_{J-\Delta+2\Delta'}}{r^{(-a,-b)}_{J+\Delta}}\frac{J+\Delta-1}{J+\Delta'-1}c^t(\Delta',J-\Delta+\Delta')
\\ & = c^t(\Delta,J)\,.
\ea
This confirms that eq.~\eqref{Regge dDisc} is precisely dual to the Lorentzian inversion formula,
at least in $d=2,4$. It is what the ``inversion'' \eqref{LIF} inverts!

\section{The Lorentzian fishnet model at subleading powers \label{fishnetsection}}

The solvability of the fishnet theory provides a testing ground for the results of the preceeding section. On the one hand, the Euclidean solution (\ref{expansion}) can be analytically continued to the Lorentzian regime directly. A nice technique for determining this continuation based on properties of HPL functions is outlined in Appendix \ref{app:HPL}. On the other hand, the correlator (\ref{corr}) is already written in the spectral decomposition required for the analytic continuation of the conformal blocks. The only challenge after applying the continuation is the navigation of complex $j$ and $\nu$ planes as the integration contours are deformed.

\subsection{The zero-magnon correlator: $u$-channel ladders}

The relevant equations for the zero-magnon correlator were outlined in section \ref{fishnetreview}. We reintroduce the shadow block to \eqref{corr} by exploiting the shadow symmetry and compare with equation \eqref{spectral}. This allows us to extract all the OPE data which can be inserted directly into equation \eqref{main} for the discontinuity. Since the $u$- and $t$-channel ladders were computed separately, we treat them separately in this section as well.  The $u$-channel ladder contributes only to the
$c^u$ part of eq.~\eqref{main} due to the $(-1)^J$ factor of eq.~\eqref{corr}.

\begin{figure}[t]
	\def\svgwidth{\linewidth}
	\centering{
		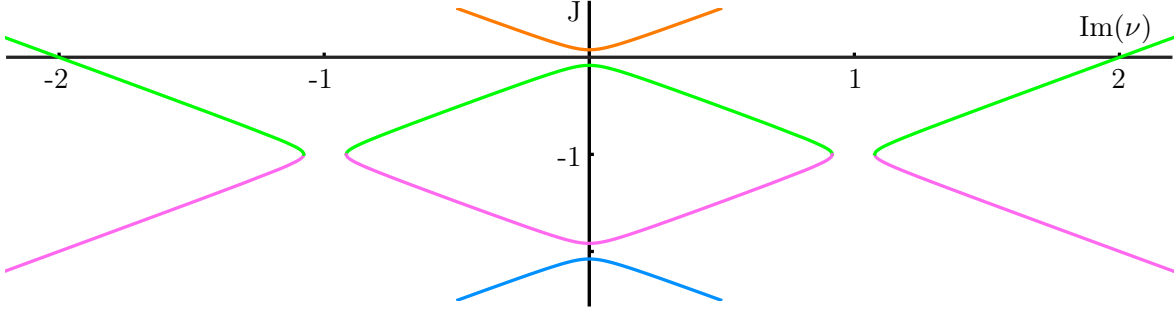
		\caption{The Regge trajectories in conformal fishnet theory. At zero coupling the trajectories collide on integer spin and scaling dimension. The four colours correspond to the solutions in (\ref{Jsolutions}), from top along vertical axis: $J_1$ is orange, $J_2$ is green, $J_3$ is pink and $J_4$ is blue. Note that $\text{Im}(\nu) = -(\Delta-2)$.}
		\label{trajectories}
	}
\end{figure}

To compute the discontinuity of the $u$-channel ladders (eq.~\eqref{corr}), we first focus on the modified
block $G^{(0,0)}_{1-\Delta,1-j}$ which enters eq.~\eqref{main}.
As prescribed, we isolate the physical poles in the $j$-plane from the four solutions to eq. (\ref{const}),
\begin{equation}\label{Jsolutions}
J(\nu, \xi)=-1\pm\sqrt{1-\nu^2\pm\sqrt{-\nu^2+4\xi^4}},
\end{equation} which are labelled $J_i$ for $i=1,\dots,4$, illustrated in figure \ref{trajectories}. These solutions correspond to the $j_n(\Delta)$'s discussed in section \ref{sommerfeld}.

%
%\begin{figure}
%	\centering
%	\includegraphics[width=0.7\linewidth]{images/trajectories}
%	\caption{$J(i \nu)$ trajectories... to be drawn properly}
%	\label{trajectories}
%\end{figure}
After evaluating the $j$-residues in eq.~\eqref{main} we only have the $\nu$ integral remaining:
\begin{align}
 \Disc_{14} \cG^u(z,\zb)= \sum_{i=1}^4 \int_{-\infty}^{\infty}d\nu \frac{1}{2 \sin(\pi J_i)}\frac{1}{J_i(2+J_i)+\nu^2} \:\Pi_{\Delta,J_i}
 \:G^{(0,0)}_{1-\Delta, 1-J_i}(z, \zb)\nonumber \\ + (F'^{(0,0)}_{\Delta, J} \textrm{ term})\,, \label{jResidue}
\end{align} where we have absorbed factors from the analytic continuation and normalization into
\begin{equation}
\Pi_{\Delta,J}= C_{\Delta,J} \frac{2^{2 (\Delta +J)}\Gamma \left(\frac{1}{2} (J+\Delta -1)\right) \Gamma \left(\frac{1}{2} (J+\Delta +1)\right)}{2(J+1) \,\Gamma\left(\frac{J+\Delta }{2}\right)^2}\,.
\end{equation}
where as before $\Delta=2+i\nu$.
%The $(-1)^J$ just cancels that hidden in $C(\nu, J)$, so this equation is consistent with the $u$-channel piece of eq.~\eqref{main}.
At lowest order in $\sigma$, only the trajectories with the positive square root, $J_1$ and $J_2$, will contribute since $G_{1-\Delta,1-j}\sim \sigma^{1-j}$. The integral over $\nu$ can be evaluated by residues, though the pole and branch structure is significantly more complicated than in the Euclidean case.  At subleading powers, all trajectories contribute.

\begin{figure}[h]
	\def\svgwidth{\linewidth}
	\centering{
		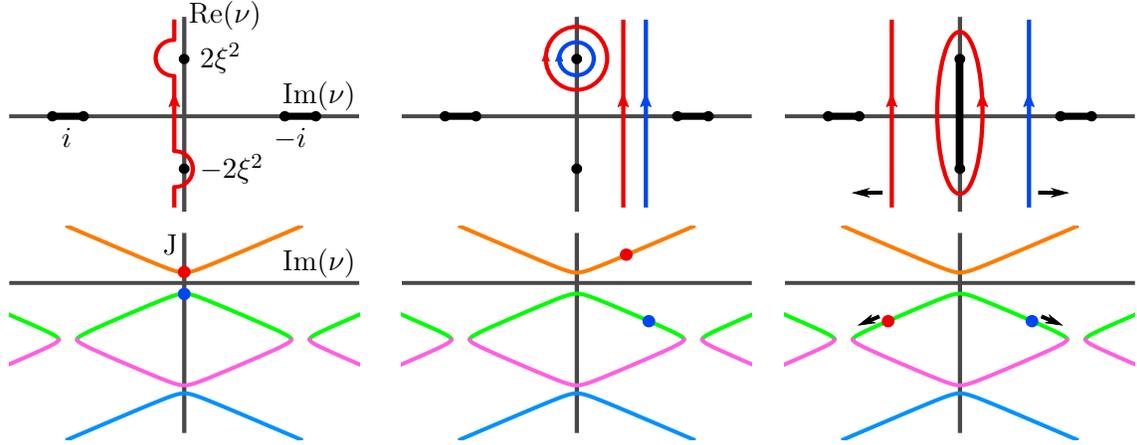
		\caption{The movement of contours along the $J$ trajectories. The upper panels indicate the $J_L$ and $J_R$ contours in the $\nu$-plane while the lower panels indicate their corresponding position on the $J$ solution sheets.
		The strategy to expand in the Regge limit is to move down as fast as possible (so as to $J$), as shown in the third column.}
		\label{Jcontour}
	}
\end{figure}

We expand the integrands of \eqref{jResidue}  to a desired order in $\sigma = z \zb$ so the $\nu$ integration becomes manageable. The initial contours for all trajectories run along the real $\nu$-axis, and for $J_1$ and $J_2$
they go below and above poles at $\nu=-2\xi^2$ and $\nu=2\xi^2$, respectively, since $\Ima(\xi^2)<0$.
The contours are then deformed as illustrated in figure \ref{Jcontour}. Each of integrands has a branch cut running between the two poles where the $J_1$ and $J_2$ sheets intersect. These cuts cancel perfectly when the two integrands are added.

The first step in the integration is to drag both contours to the right, picking up a residue at $\nu=2\xi^2$.
In the Regge limit we strive to decrease $J$ (to make the integrand as small as possible) and so
we must drag the $J_1$ contour back across its branch cut and onto the $J_2$ sheet, as drawn in panel 3 of figure \ref{Jcontour}. Since the labellings no longer refer to the original solutions, we relabel the integrands as $J_L$ on the left and $J_R$ on the right. The residues at this step contribute to $ \Disc_{14} \cG^u(z,\zb)$ at order $\sigma$ and higher. We also find that within an $\cO(\xi^4)$ radius each of the branch points and poles shown in figure \ref{Jcontour} there are poles from the cosecant function in the $\nu$-plane that must be included in the calculation. They are not included in the plots because in general we expand in small $\xi$ for calculations, at which point the cosecant poles coinside with the plotted solutions. Later on in the calculations the poles from the cosecant will be independent and must then be treated independently.

At leading power, the correlator is thus saturated by the contribution from the branch cut shown in the third column
of fig.~\ref{Jcontour}, ranging over $\nu\in [-2\xi^2,2\xi^2]$.  This phenomenon was observed in refs.~\cite{Korchemsky:2018hnb,Chowdhury:2019hns}.  At subleading powers, we will obtain similar contributions from other intersections, as we now see.

\begin{figure}[h]
	\def\svgwidth{\linewidth}
	\centering{
		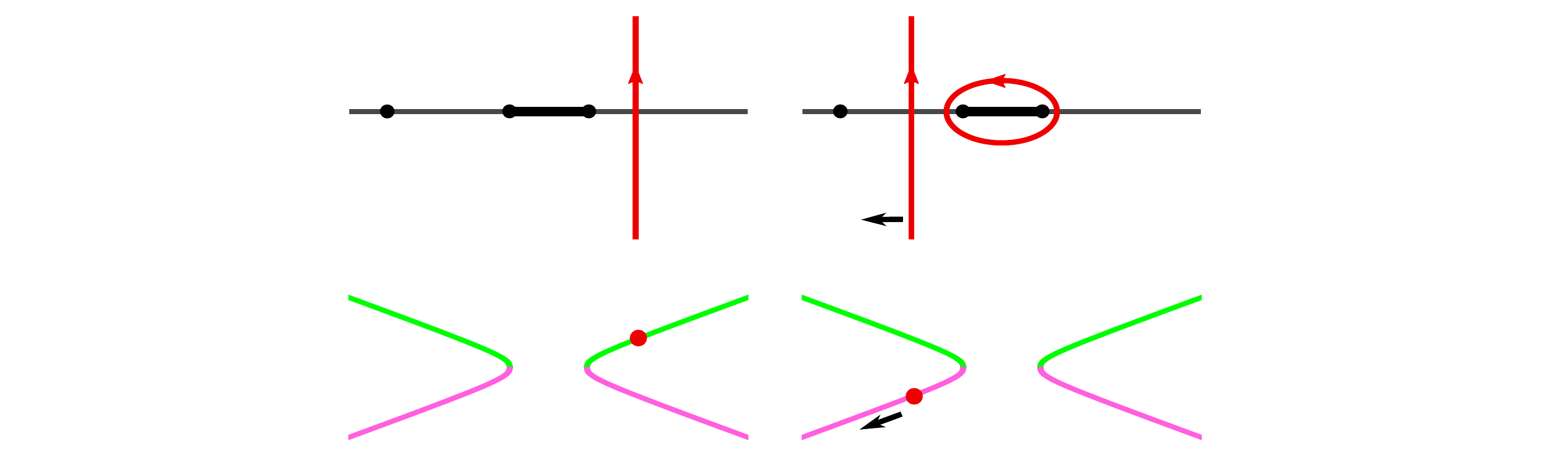
		\caption{The movement of contours along the $J_L$ trajectory past the branch cut around $\nu=i$. As the contour is dragged further to the left along the $J_3$ solution it will pick up residues of the cosecant function when $J_3(\nu)$ is an integer.}
		\label{jResidue2}
	}
\end{figure}

The next feature that the $J_L$ contour encounters is a branch cut running from $\nu=-i-2\xi^2 -\cO(\xi^4)$ to $\nu=-i+2\xi^2+\cO(\xi^4)$. This branch is analogous to the first but for the $J_2$ and $J_3$ intersection. The contribution to the integrand at this location is the contour around the branch cut, as illustrated in figure \ref{jResidue2}. The $J_L$ contour can now be dragged to imaginary infinity in the $\nu$-plane with the only obstructions being poles of the cosecant at integer $j$. The first of these is at $J_3=-2$ and hence contributes at $\sigma^3$. Due to the $\nu \leftrightarrow -\nu$ symmetry of the integrand, the computation for the right-moving contour is equivalent up to signs from contour orientations. Since $J=-1$ around these branches, these residues contribute to $ \Disc_{14} \cG^u(z,\zb)$ at order $\sigma^2$.

\begin{figure}[h]
	\def\svgwidth{\linewidth}
	\centering{
		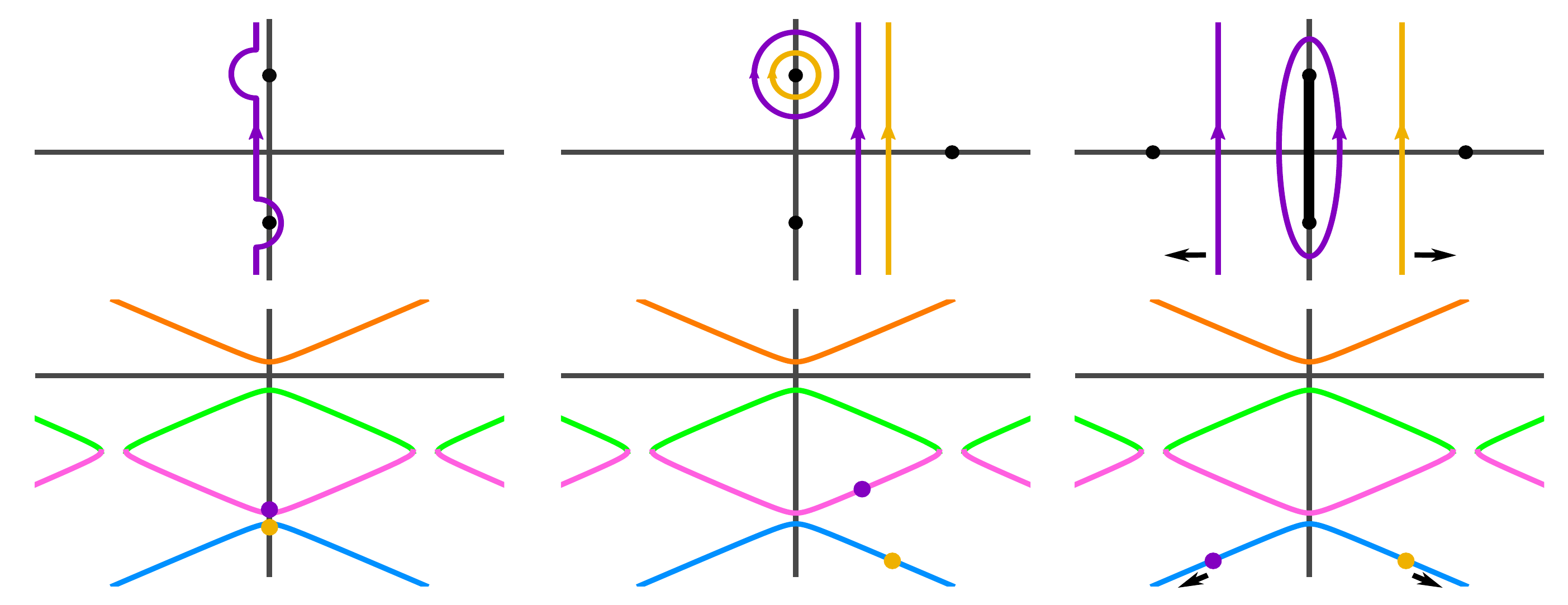
		\caption{The third intersection integration is analogous to the first. The integration contours will pick up residues at the poles of the cosecant as they are pulled to larger negative $J$.}
		\label{jResidue3}
	}
\end{figure}

\begin{figure}[h!]
	\def\svgwidth{1\linewidth}
	\centering{
		%% Creator: Inkscape 1.0 (4035a4f, 2020-05-01), www.inkscape.org
%% Accompanies image file 'contributions2.pdf' (pdf, eps, ps)
\begingroup%
  \makeatletter%
  \providecommand\color[2][]{%
    \errmessage{(Inkscape) Color is used for the text in Inkscape, but the package 'color.sty' is not loaded}%
    \renewcommand\color[2][]{}%
  }%
  \providecommand\transparent[1]{%
    \errmessage{(Inkscape) Transparency is used (non-zero) for the text in Inkscape, but the package 'transparent.sty' is not loaded}%
    \renewcommand\transparent[1]{}%
  }%
  \providecommand\rotatebox[2]{#2}%
  \newcommand*\fsize{\dimexpr\f@size pt\relax}%
  \newcommand*\lineheight[1]{\fontsize{\fsize}{#1\fsize}\selectfont}%
  \ifx\svgwidth\undefined%
    \setlength{\unitlength}{779.52755906bp}%
    \ifx\svgscale\undefined%
      \relax%
    \else%
      \setlength{\unitlength}{\unitlength * \real{\svgscale}}%
    \fi%
  \else%
    \setlength{\unitlength}{\svgwidth}%
  \fi%
  \global\let\svgwidth\undefined%
  \global\let\svgscale\undefined%
  \makeatother%
  \begin{picture}(1,0.3881313)%
    \lineheight{1}%
    \setlength\tabcolsep{0pt}%
    \put(0,0){\includegraphics[width=\unitlength,page=1]{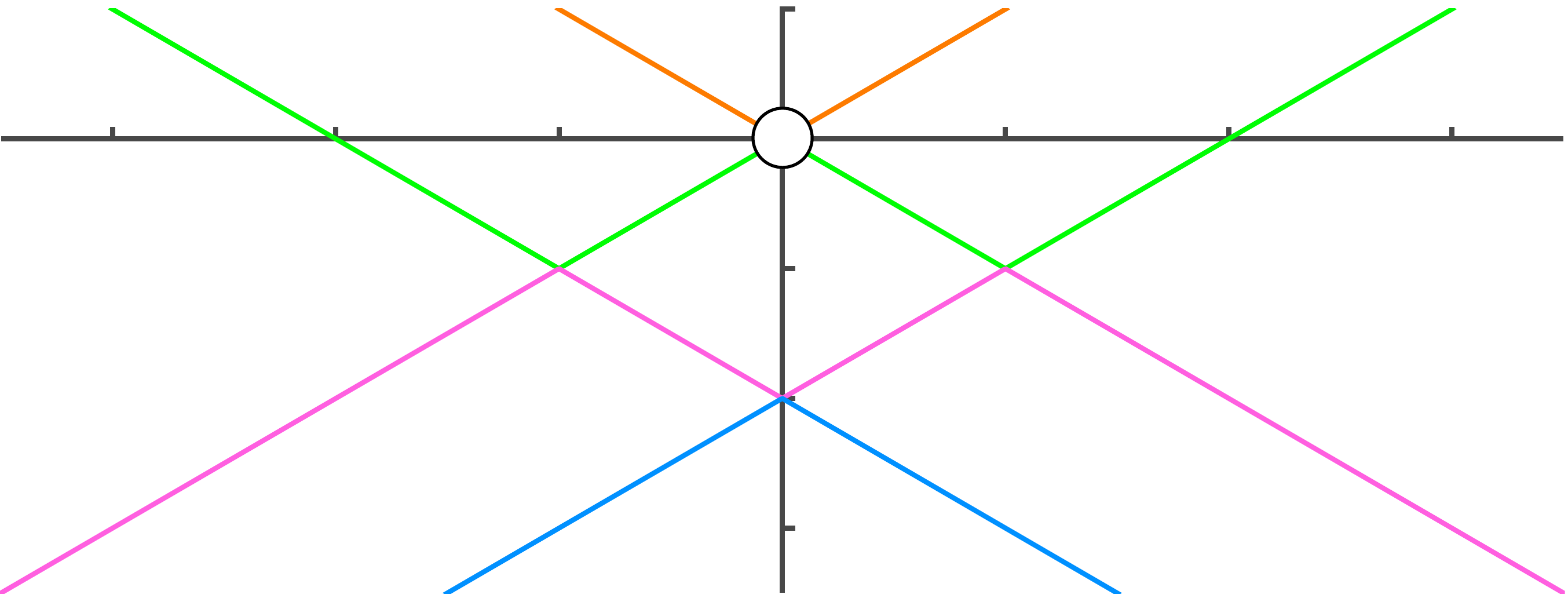}}%
    \put(0.48862441,0.29310149){\color[rgb]{0,0,0}\makebox(0,0)[lt]{\lineheight{1.25}\smash{\begin{tabular}[t]{l}$A$\end{tabular}}}}%
    \put(0,0){\includegraphics[width=\unitlength,page=2]{contributions2.pdf}}%
    \put(0.48958658,0.12598821){\color[rgb]{0,0,0}\makebox(0,0)[lt]{\lineheight{1.25}\smash{\begin{tabular}[t]{l}$C$\end{tabular}}}}%
    \put(0,0){\includegraphics[width=\unitlength,page=3]{contributions2.pdf}}%
    \put(0.633278,0.20707283){\color[rgb]{0,0,0}\makebox(0,0)[lt]{\lineheight{1.25}\smash{\begin{tabular}[t]{l}$B$\end{tabular}}}}%
    \put(0,0){\includegraphics[width=\unitlength,page=4]{contributions2.pdf}}%
    \put(0.34614453,0.20761522){\color[rgb]{0,0,0}\makebox(0,0)[lt]{\lineheight{1.25}\smash{\begin{tabular}[t]{l}$B$\end{tabular}}}}%
    \put(0,0){\includegraphics[width=\unitlength,page=5]{contributions2.pdf}}%
  \end{picture}%
\endgroup%

		\caption{Contributions to the Regge-limit four-point correlator of Conformal Fishnet Theory. Cuts $A$-$C$ occur at the intersections of the trajectories and the calculations correspond to figures \ref{Jcontour}, \ref{jResidue2} and \ref{jResidue3}. The additional maroon dots occur at the poles of the cosecant function. As hoped, all terms come from localized points or short cuts on the Regge trajectories. }
		\label{contributions}
	}
\end{figure}

Also at order $\sigma^3$, we must consider the $J_3$ and $J_4$ trajectories at $J=-2$ in the same way as the $J=0$ intersection. This contour deformation is illustrated in figure \ref{jResidue3}. Additional contributions to the correlation function at $\sigma^4$ come from the poles of the cosecant function along the $J_3$ and $J_4$ trajectories, arising from all the contours being pulled to more negative $J$. These additional points are plotted in fig. \ref{contributions}.

To summarize, the $\nu$ integration contour was deformed along the Regge trajectories while collecting contours around poles and branches at the $J=0$ intersection which start contributing to $ \Disc_{14} \cG^u(z,\zb)$ at order $\sigma$, branch contours at $J=-1$ of order $\sigma^2$ and pole and branch contours at $J=-2$ of order $\sigma^3$.

The $F'^{(0,0)}_{J,\Delta}(z,\zb)$ term is the remaining contributor to the discontinuity \eqref{jResidue}. It can be evaluated in the same manner as the Euclidean four-point function but with an additional coefficient from the $s_\beta^{(0,0)}$ term in eq.~\eqref{main}, which in this case reduces to
\begin{equation}
G^{(0,0)}_{\Delta, J}\to -\tan\Big(\frac{\pi}{2} (J+\Delta)\Big)G^{(0,0)}_{\Delta, J}\,.
\end{equation}

We can now present the various contour integration results up to $\sigma^3$ and to the first two orders of $\xi^2$. The corresponding locations on the $\nu$-contours are indicated in figure \ref{contributions}. We computed these terms along with additional residues at the poles of the cosecant function up to orders ($\sigma^4$, $\xi^8$) and ($\sigma^3$, $\xi^{12}$).
We found perfect agreement with the direct HPL continuations; to illustrate the nontrivial interplay between the contributions, we now record explicit formulas at lower order.

\noindent
The first intersection ($\nu=0$, $J=0$) contributes:
\begin{multline}
	iA =\sigma  \left(-\frac{4 \pi  \xi ^2 w \log (w)}{(w-1) (w+1)}-\frac{4 i \pi  \xi ^4 w (\log (\sigma )-1) \log (w)}{(w-1) (w+1)} \right) + \\ \sigma ^2 \left(- \pi  \xi ^2-\frac{ i \pi  \xi ^4 \left(-\log (\sigma )+w^2 \log (\sigma )-w^2+w^2 \log (w)+\log (w)+1\right)}{(w-1) (w+1)}\right)+\\\sigma ^3 \left(-\frac{\pi  \xi ^2 \left(w^2+1\right)}{2 w}-\frac{i \pi  \xi ^4 \left(-\log (\sigma )+w^4 \log (\sigma )-w^4+w^4 \log (w)+\log (w)+1\right)}{2 (w-1) w (w+1)}\right)\\ +\cO(\sigma^4, \xi^6)\,.
\end{multline}
The second intersection ($\nu=\pm i$, $J=-1$):
\begin{multline}
	iB = \sigma ^2\left(\frac{ i \pi  \xi ^4  \left(\log (\sigma )-w^2 \log (\sigma )+w^2+w^2 \log (w)+\log (w)-1\right)}{(w-1) (w+1)}\right) + \\\sigma ^3\left( \frac{i \pi  \xi ^4  \left(\log (\sigma )-w^4 \log (\sigma )+w^4+w^4 \log (w)+\log (w)-1\right)}{2 (w-1) w (w+1)}\right)\\ +\cO(\sigma^4, \xi^6)\,.
\end{multline}
The third Intersection ($\nu = 0$, $J=-2$):
\begin{equation}
	iC = \sigma ^3 \left(\frac{ \pi  \xi ^6 w \log (w)}{(w-1) (w+1)}+\frac{ i \pi  \xi ^8 w (\log (\sigma )-2) \log (w)}{(w-1) (w+1)}\right) +\cO(\sigma^4, \xi^{10})\,.
\end{equation}
The cosecant poles ($\nu = \pm 2 i$, $J=-2$):
\begin{equation}
	iD = \sigma ^3\left(-\frac{i \pi  \xi ^4 \left(w^2+1\right)}{4 w}-\frac{i \pi  \xi ^8 \left(-11 w^4+4 w^4 \log (w)+4 \log (w)+11\right)}{16 (w-1) w (w+1)}\right)\\ +\cO(\sigma^4, \xi^{10})\,.
\end{equation}
Finally, the $F'^{(0,0)}_{J,\Delta}(z,\bar{z})$ contribution:
\begin{multline}
	iE = \sigma ^2 \left(\pi  \xi ^2+2 i \pi  \xi ^4 (\log (\sigma )-1)\right)+\\\sigma ^3 \left(\frac{\pi  \xi ^2 \left(w^2+1\right)}{2 w}+\frac{i \pi  \xi ^4 \left(w^2+1\right) (4 \log (\sigma )-3)}{4 w}\right)+\cO(\sigma^4, \xi^6)\,.
\end{multline}
\noindent
The sum of these expressions $A-E$ is then found to be
equal to the $\Disc_{14}$ of the fishnet correlation function given in equation (\ref{expansion}),
after the analytic continuation of harmonic polylogarithms detailed in appendix \ref{app:HPL}:
\begin{multline}
\Disc_{14}\cG^u(z,\bar{z}) = \sigma  \bigg(\frac{4 i \pi  \xi ^2 w \log (w)}{(w-1) (w+1)}-\frac{4 \pi  \xi ^4 w (\log (\sigma )-1) \log (w)}{(w-1) (w+1)}\\-\frac{2 i \pi  \xi ^6 w \log (w) \left(4 \log ^2(w)+2 \pi ^2-9\right)}{3 (w-1) (w+1)}\bigg)+\frac{1 }{3}\sigma ^2 \xi ^6  (-4 \log ^3(\sigma )+6 i \pi  \log ^2(\sigma )+\\12 \log ^2(\sigma )-12 i \pi  \log (\sigma )-15 \log (\sigma )-6 i \pi  \log ^2(w)-12 \zeta (3)+12 i \pi +15) + \cO(\sigma^3,\xi^8)\,.
\end{multline}

\subsection{$t$-channel ladders and their double discontinuity \label{tchannel}}

The Regge limit can also be considered for the $t$-channel ladders, for which the Euclidean OPE is given by the same expression as eq.~\eqref{corr} but without the overall $(-1)^J$ factor.
The $t$-channel data is interesting because it is the only contributor to $\dDisc\cG(z,\zb)$ (see eq.~\eqref{Regge dDisc}).
We checked that the Sommerfeld-Watson calculation matched the direct HPL continuation up to order $\sigma^2$ and $\xi^8$, this time including the $e^{i\pi J}$ factor in eq. \eqref{main}. At this level, we would have detected any errors in the formula for the analytic continuation that would not have been sensed in the $u$-channel case. The $t$-channel HPL continutations were obtained from the $u$-channel results by substituting $z\to z/(z-1)$ and $\zb\to \zb/(\zb-1)$ with appropriate phases.

In section \ref{confregge} we presented equation \eqref{Regge dDisc} for the double discontinuity in terms of a double integral of the OPE data and the Regge block over spin and scaling dimension, which we also checked explicitly using the $t$-channel ladders the fishnet model. The calculations are performed almost identically to those for the $\Disc_{14}$, however the cosecant function in spin is removed along with several constant coefficients from definitions. For the $G_{1-\Delta, 1-J}$ block, the analytic structure and $\nu$ integration follows the contour deformations drawn in figures \ref{Jcontour}, \ref{jResidue2} and \ref{jResidue3}. The contributions from the poles at $\nu=\pm 2\xi^2$ vanish and therefore the double discontinuity contains only terms at even powers of $\xi^2$. Moreover, there are no longer poles from a cosecant contributing at orders $\sigma^3$ and higher. In reference to fig. \ref{contributions}, only terms from locations $A$-$C$ contribute. As before, the remaining shadow blocks are irrelevant for the calculation.

We verified that equation \eqref{Regge dDisc} was correct for the zero-magnon four-point function by comparing the direct integration of the \eqref{Regge dDisc} to the analytic continuations of the HPL functions appearing in the first line of \eqref{dDisc def}. We computed the double discontinuity to orders ($\sigma^4, \xi^{12}$) in both ways and found perfect agreement. In the Regge limit, the double discontinuity is
\begin{multline}
\dDisc[ \cG(z,\bar{z})] = \sigma  \left(\frac{4 \pi ^2 \xi ^4 w \log (w)}{(w-1) (w+1)}-\frac{2 \pi ^2 \xi ^8 w \log (w) (-3 \log ^2(\sigma )+3 \log ^2(w)+\pi ^2)}{3 (w-1) (w+1)}\right)\\+\sigma ^2\frac{2 \pi ^2 \xi ^8}{(w-1) (w+1)}  \Big(-\log ^2(\sigma )+\log (\sigma )+w^2 \log ^2(\sigma )-w^2 \log (\sigma )\\-w^2 \log ^2(w)+w^2 \log (w)+\log ^2(w)+\log (w)\Big)+\cO(\sigma^3, \xi^{12})\,.
\end{multline}

\subsection{The one-magnon correlator \label{onemagnon}}

We can follow the main steps of the zero-magnon case to compute the Regge limit of the one-magnon four-point function. The interest is that the external operators have varying scaling dimensions, namely $\Delta_1=\Delta_4=2$ and $\Delta_2=\Delta_3=1$, so it will allow us to further verify the equations of section \ref{confregge} ($a$ and $b$ vanished in the zero-magnon case and are now non-zero, $a=b=-1/2$). Interestingly, the Regge trajectories are significantly simpler! Moreover, when computing contributions to the Regge limit, there is only one branch cut to worry about (which contributes at leading order) and the only additional features leading to subleading corrections include the poles of the cosecant function attached to the $G_{1-J,1-\Delta}$ block and sum over spins in the $F'_{\Delta, J}$ block. Since the formulae for the analytic continuations are relevant at the leading order ($\sigma^2$ in this case) and first subleading order (now $\sigma^3$), we compute $\Disc_{14}$ of the one-magnon correlator only to order $\sigma^3$ and $\xi^6$.  We again find agreement with known fishnet data once analytically continued to the Lorentzian regime and evaluated at high energy.

\begin{figure}[t]
	\centering{
		\def\svgwidth{0.7\linewidth}%% Creator: Inkscape 1.0 (4035a4f, 2020-05-01), www.inkscape.org
%% Accompanies image file 'FishnetDiagram1Magnon.pdf' (pdf, eps, ps)
\begingroup%
  \makeatletter%
  \providecommand\color[2][]{%
    \errmessage{(Inkscape) Color is used for the text in Inkscape, but the package 'color.sty' is not loaded}%
    \renewcommand\color[2][]{}%
  }%
  \providecommand\transparent[1]{%
    \errmessage{(Inkscape) Transparency is used (non-zero) for the text in Inkscape, but the package 'transparent.sty' is not loaded}%
    \renewcommand\transparent[1]{}%
  }%
  \providecommand\rotatebox[2]{#2}%
  \newcommand*\fsize{\dimexpr\f@size pt\relax}%
  \newcommand*\lineheight[1]{\fontsize{\fsize}{#1\fsize}\selectfont}%
  \ifx\svgwidth\undefined%
    \setlength{\unitlength}{510.23622047bp}%
    \ifx\svgscale\undefined%
      \relax%
    \else%
      \setlength{\unitlength}{\unitlength * \real{\svgscale}}%
    \fi%
  \else%
    \setlength{\unitlength}{\svgwidth}%
  \fi%
  \global\let\svgwidth\undefined%
  \global\let\svgscale\undefined%
  \makeatother%
  \begin{picture}(1,0.30555556)%
    \lineheight{1}%
    \setlength\tabcolsep{0pt}%
    \put(0,0){\includegraphics[width=\unitlength,page=1]{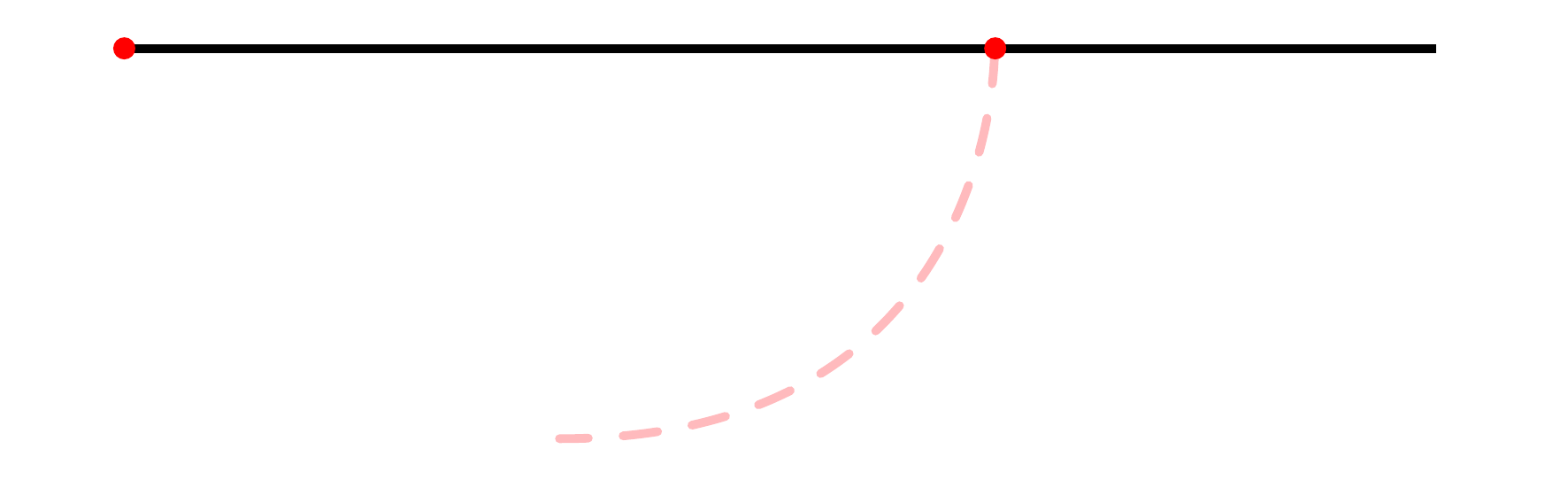}}%
    \put(0.02795796,0.24410214){\color[rgb]{0,0,0}\makebox(0,0)[lt]{\lineheight{1.25}\smash{\begin{tabular}[t]{l}$x_1$\end{tabular}}}}%
    \put(0.02795796,0.04464654){\color[rgb]{0,0,0.07843137}\makebox(0,0)[lt]{\lineheight{1.25}\smash{\begin{tabular}[t]{l}$x_2$\end{tabular}}}}%
    \put(0.93138114,0.24410214){\color[rgb]{0,0,0}\makebox(0,0)[lt]{\lineheight{1.25}\smash{\begin{tabular}[t]{l}$x_3$\end{tabular}}}}%
    \put(0.93138114,0.04464654){\color[rgb]{0,0,0.07843137}\makebox(0,0)[lt]{\lineheight{1.25}\smash{\begin{tabular}[t]{l}$x_4$\end{tabular}}}}%
    \put(0,0){\includegraphics[width=\unitlength,page=2]{FishnetDiagram1Magnon.pdf}}%
  \end{picture}%
\endgroup%

		\caption{An example of a fishnet ladder diagram evaluated in the computation of $\cG(z,\zb)$ with one-magnon operators. The four-point interaction sites have coupling $\xi^2$ so this diagram contributes at order $\xi^{4}$ and the shading of the propagator indicates how it ``winds''. The points $x_1$ and $x_2$ ``source'' the propagator and account for the modified scaling dimensions of the operators inserted at these points (see \cite{Gromov:2018hut} for additional diagrams and discussion).}
		\label{fishnetmagnon}
	}
\end{figure}

We first review the physics of the Euclidean one-magnon four-point function derived in \cite{Gromov:2018hut} and sketched in fig. \ref{fishnetmagnon}. The correlator is given by (again, trace implied)
\be
\<0 | Y(x_1)X(x_1)Y(x_2)Y^\dagger(x_3)Y^\dagger(x_4)X^\dagger(x_4)|0\> \equiv \frac{(x_{13}^2x_{24}^2)^{1/2}}{x^2_{14}(x_{12}^2x_{34}^2)^{3/2}} \cG_1(z,\zb)
\ee
The sum over conformal blocks is slightly modified to
\begin{equation}\label{corrtwo}
\cG_1(z,\zb)=\sum_{J\geq 0}  \int_{-\infty}^{\infty} \frac{d\nu}{2\pi} C^\prime_{\Delta,J} \frac{ 256(-1)^JH^2_{\Delta,J}}{1-16\: \xi^2 H_{\Delta,J} } \: G^{(-\frac{1}{2},-\frac{1}{2})}_{\Delta,J}(z,\zb),
\end{equation}
with a new set of energy eigenvalues,
\be\label{energyeqntwo}
H_{\Delta,J}=\frac{(-1)^J}{4(-\Delta+J+3)(\Delta+J-1)}\,.
\ee
A novelty is that $(-1)^J$ occurs both in the numerator and denominator of eq.~\eqref{corrtwo};
as remarked in \cite{Gromov:2018hut}, its appearance in $H$ is necessary to cancel the spurious poles of the blocks with the spurious poles of the normalization coefficient, which in this case includes the external operator dimensions \cite{Gromov:2018hut}
\begin{multline}
C^\prime_{\Delta,J} =  \frac{\Gamma(\Delta-1)\Gamma(2+J) \Gamma(4-\Delta+J)}{2\Gamma(J+1)\Gamma(\Delta-2) \Gamma(\Delta+J-1)}\\
\times \frac{\Gamma(\frac{1}{2}(\Delta+J+\Delta_1-\Delta_2))\Gamma(\frac{1}{2}(\Delta+J-\Delta_1+\Delta_2))}
{\Gamma(2-\frac{1}{2}(\Delta-J-\Delta_1+\Delta_2))\Gamma(2-\frac{1}{2}(\Delta-J+\Delta_1-\Delta_2))}\,.
\end{multline}
Just as in the zero-magnon case,  the correlator can be expanded in the coupling,
\begin{equation}\label{expansiontwo}
\cG_1(z,\zb)=\frac{(z \zb)^{3/2}}{z-\zb}\sum_{n=0}^{\infty} \xi^{2n} \cG_1^{(n)}(z,\zb)\,,
\end{equation}
and the $\cG_1^{(n)}$'s are combinations of HPLs. At leading order we have
\begin{equation}
\cG_1^{(0)}(z,\zb) = z-\zb\,.
\end{equation} We verified the expansions from \cite{Gromov:2018hut} to to order $\sigma^4$ and $\xi^8$.

\begin{figure}[t]
	\def\svgwidth{\linewidth}
	\centering{
		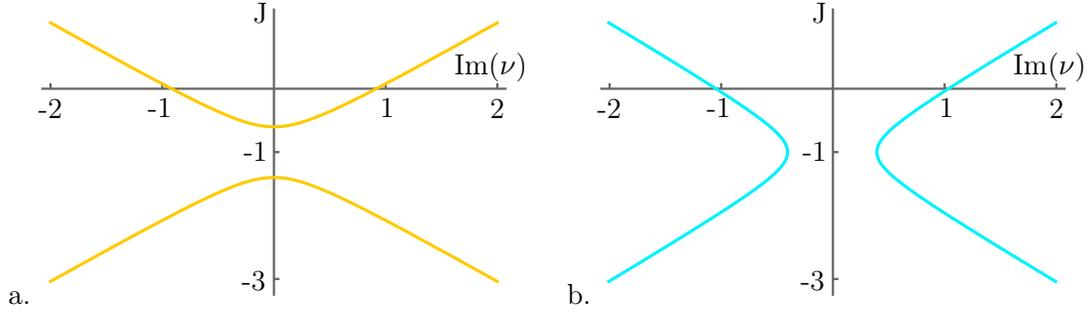
		\caption{Regge trajectories for the one-magnon four-point function. The $J^{\rm even}$ and $J^{\rm odd}$ trajectories are in the left and right panels, respectively. Note that the gap between the curves has length of order $\xi^2$.}
		\label{1magtraj}
	}
\end{figure}

\begin{figure}[h]
	\def\svgwidth{\linewidth}
	\centering{
		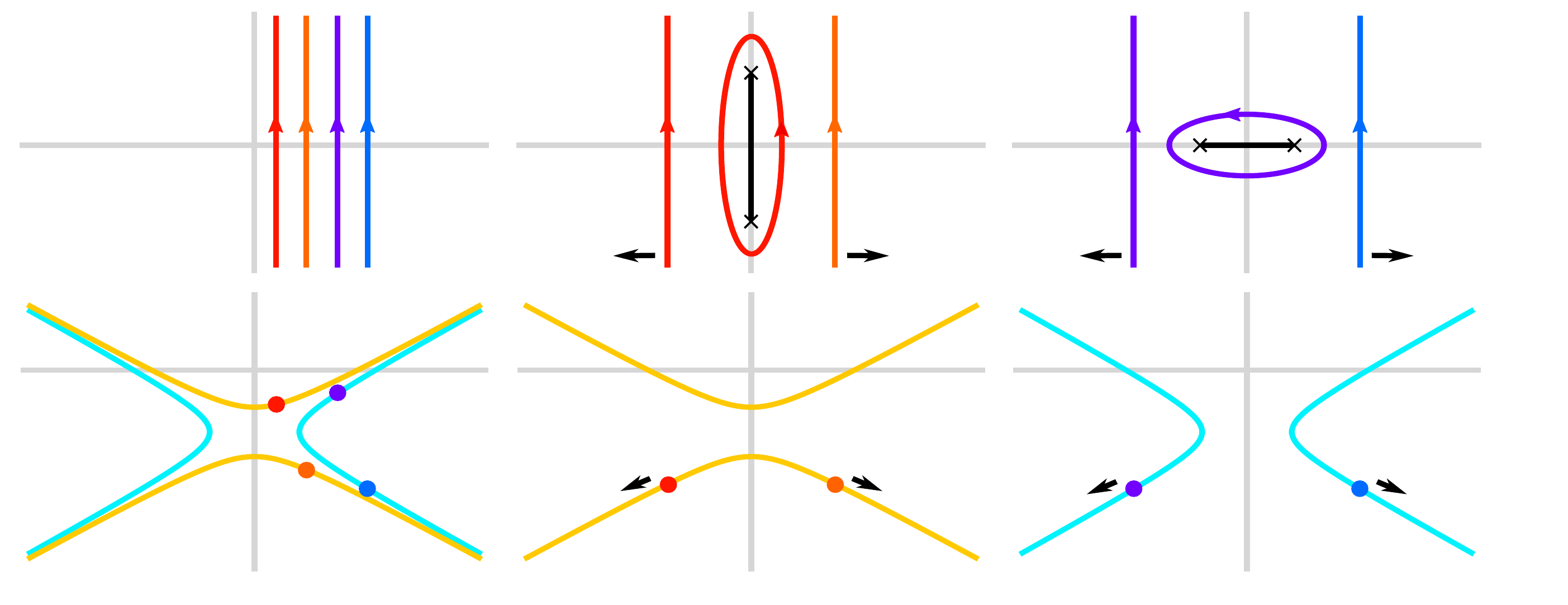
		\caption{A plot of the $\nu$ integration along the Regge trajectories, following fig.~\ref{Jcontour}.
		The warm and cool colours correspond to the $J^{\rm even}$ and $J^{\rm odd}$ solutions, respectively.}
		\label{1magintersection}
	}
\end{figure}

We now wish to analytically continue our correlator (\ref{corrtwo}) to the Lorentzian kinematics regime. To make contact with eq. \eqref{partialwavesplitting}, one can work out that
\begin{align}\label{OPEdata}
c^t(\Delta, J)&=\frac{1}{2}(c^{\rm even}(\Delta, J) +c^{\rm odd}(\Delta, J))\nonumber \\ 
c^u(\Delta, J)&=\frac{1}{2}(c^{\rm even}(\Delta, J) - c^{\rm odd}(\Delta, J))\,,
\end{align}
 where $c^{\rm even}$ and $c^{\rm odd}$ represent the OPE data of the correlator \eqref{corrtwo} with even and odd spin, respectively, that is, with $(-1)^J$ set to $\pm1$.
 Unlike the zero-magnon case, both channels contribute to the discontinuity $\Disc_{14}\cG_1$.
The Regge trajectories were computed by solving for the physial poles of the correlation function and are plotted in fig. \ref{1magtraj}. We denote them
\be
J^{\rm even}_\pm(\nu, \xi)= -1\pm\sqrt{-\nu^2+4\xi^2}\,\,\,\textrm{ and }\,\,\,J^{\rm odd}_\pm(\nu, \xi) = -1\pm\sqrt{-\nu^2-4\xi^2}\,.
\ee
We can now plug our OPE data into our main equation \eqref{main}, in which our Regge trajectories take the place of the $j_n(\nu)$'s. We have
\begin{align}
\Disc_{14}\cG_1 = \int_{-\infty}^{\infty}d\nu \sum_{j_n(\nu)}\Res_{J=j_n(\nu)}\frac{1}{\sin(\pi J)}\,\Pi_1(\Delta,J)\, G^{(-\frac{1}{2},-\frac{1}{2})}_{1-\Delta,1-J}(z,\bar{z})\nonumber \\ + \,(F'^{(-\frac{1}{2},-\frac{1}{2})}(z,\bar{z})\textrm{ term})\,,
\end{align}
where we collected the OPE data into
\begin{multline}
\Pi_1(\Delta, J) = C^\prime_{\Delta,J}\frac{2^{2 \Delta +2 J}\, \Gamma (\frac{J}{2}+\frac{\Delta }{2})^2}{2(-\Delta +J+3)\Gamma (\frac{J}{2}+\frac{\Delta }{2}+\frac{1}{2})^2} \\
\times \frac{(4 e^{i \pi  J} \xi ^2+(J+1)^2+\nu ^2)}{ \big( (J+1)^2+\nu^2-4 \xi ^2\big) \big( (J+1)^2+\nu^2+4 \xi ^2\big)}\,.
\end{multline}
The $e^{i\pi J}$ clearly distinguishes the $t$- and $u$-channel data. The $\nu$ integration proceeds very much like the zero-magnon case. Since the Regge intercepts is now $J=-1$ (see fig. \ref{1magtraj}), the leading order physics comes at order $\sigma^{1-J_{\rm max}}=\sigma^2$. The analytic structure at $J=-1$ is similar to the zero-magnon cases except that instead of poles at the end of the branch cuts, we find only branch points. Thus the contribution at leading order comes only from the residue around the branch cut. For the $J^{\rm even}$ trajectories, this branch runs from $-2\xi^2$ to $2\xi^2$ and the integration contours are deformed similarly to fig. \ref{Jcontour}. For the $J^{\rm odd}$ trajectories, the branch runs from  $-2i\xi^2$ to $2i\xi^2$ and the contours look more like those in fig. \ref{jResidue2}. The movement of these contours is plotted in fig. \ref{1magintersection}. 

At subleading orders in $\sigma$ we have to take into account the poles of the cosecant function. Only the integrands involving the $J^{\rm even}$ solution contribute at this order.

Additional subleading terms come from the block whose integration contour wasn't deformed. As in the zero-magnon case, we simply have to add a coefficient to the integrand of the Euclidean case to compute the $F'^{(-\frac{1}{2},-\frac{1}{2})}(z,\bar{z})$ contribution, which becomes
\be
G^{(-\frac{1}{2},-\frac{1}{2})}_{\Delta, J}\to -\cot\Big(\frac{\pi}{2}(J+\Delta)\Big)G^{(-\frac{1}{2},-\frac{1}{2})}_{\Delta, J}\,.
\ee

We can now present the results of the one-magnon calculations at order $\sigma^3$ and the first two orders of $\xi$. We checked to orders $(\sigma^3$, $\xi^6$) that these contributions matched the direct HPL continuations.

\noindent
The first intersection along $J^{\rm even}$ ($\nu=0$, $J=-1$):
\begin{multline}
iA^{\rm even} = \sigma ^2 \bigg(\frac{\pi ^2 \xi ^4 w (2 \log (\sigma )-i \pi ) \log (w)}{(w-1) (w+1)}+\\\frac{\pi ^2 \xi ^6 w \log (w) \left(4 \log ^3(\sigma )-6 i \pi  \log ^2(\sigma )-4 \log (\sigma ) \log ^2(w)+2 i \pi  \log ^2(w)-24 \zeta (3)-i \pi ^3\right)}{6 (w-1) (w+1)}\bigg)+\\\sigma ^3 \bigg(\pi ^2 \xi ^4+\frac{\pi ^2 \xi ^6}{(w-1) (w+1)} (-\log ^2(\sigma )+i \pi  \log (\sigma )+2 \log (\sigma w )+w^2 \log ^2(\sigma )-i \pi  w^2 \log (\sigma )\\-2 w^2 \log (\sigma/w )+i \pi  w^2-w^2 \log ^2(w)+\log ^2(w)-i \pi )\bigg) + \cO(\sigma^4, \xi^8)\,.
\end{multline}
The first intersection along $J^{\rm odd}$ ($\nu=0$, $J=-1$):
\begin{multline}
iA^{\rm odd}=\sigma ^2 \Big(\frac{4 i \pi  \xi ^2 w \log (w)}{(w-1) (w+1)}-\frac{\pi  \xi ^4 w \log (w) \left(2 i \log ^2(\sigma )+2 \pi  \log (\sigma )-2 i \log ^2(w)-i \pi ^2\right)}{(w-1) (w+1)}\Big)\\-\sigma ^3 \frac{\pi  \xi ^4 \left(-2 i \log (\sigma )+2 i w^2 \log (\sigma )+\pi  w^2-2 i w^2 \log (w)-2 i \log (w)-\pi \right)}{(w-1) (w+1)}+\cO(\sigma^4, \xi^6)\,.
\end{multline}
The cosecant poles (only from $J^{\rm even}$) ( $\nu=\pm i$, $J=-2$):
\begin{equation}
iB = \sigma ^3 \bigg(i \pi  \xi ^2-\frac{i \pi  \xi ^4 \left(-3 w^2+2 w^2 \log (w)+2 \log (w)+3\right)}{(w-1) (w+1)}\bigg) + \cO(\sigma^4, \xi^6)\,.
\end{equation}
The $F'^{(-\frac{1}{2},-\frac{1}{2})}(z,\bar{z})$ contribution:
\begin{equation}
iC = \sigma ^3 \left(-i \pi  \xi ^2+i \pi  \xi ^4 (2 \log (\sigma )-3)\right)+ \cO(\sigma^4, \xi^6)\,.
\end{equation}
The sum of these results match the continued HPLs! The full discontinuity in the Regge limit is
\begin{multline}
\Disc_{14}\cG_1(z,\bar{z}) = \sigma ^2 \bigg(\frac{4 \pi  \xi ^2 w \log (w)}{(w-1) (w+1)}+\frac{2 \pi  \xi ^4 w \log (w) (\log (w)-\log (\sigma )) (\log (\sigma )+\log (w))}{(w-1) (w+1)}\bigg)\\+\cO(\sigma^4, \xi^6)\,.
\end{multline}

\section{Conclusion} \label{sec:conclusion}

This paper extended the formalism of Regge theory applied to four-point correlation functions in conformal field theories.
Our main result, eq.~\eqref{Regge disc} provides an exact expression for the resummed OPE in a Lorentzian spacetime, which can be expanded at high energies according to eq.~\eqref{main} to compute subleading power corrections
in a given model.
At leading power, the formula reproduces existing work from the conformal bootstrap literature.
The key new ingredient is the Regge block $R$ defined in eq.~\eqref{Rab}, which allows to seamlessly deal with subleading powers.  We also obtained an exact representation for the expectation value of a double commutator,
eq.~\eqref{Regge dDisc}.

The second goal of this paper was to check eq.~\eqref{main} explicitly in conformal fishnet theory, a treasure trove of data. We found perfect agreement to high orders in energy and the coupling in both the zero- and one-magnon four-point functions. 

As mentioned in introduction, we expect this formula to be useful in situations which require
going beyond the single-exchange approximation, such as situations
involving saturation or for precision studies in theories where forward scattering is
asymptotically transparent.

%There are several applications to the formulae derived in this paper. For instance, it is well known that the Regge limit coincides with the chaos regime in two dimensional CFTs \cite{Roberts_2015}. Therefore, in chaotic theories such as the two-dimensional supersymmetric SYK model, one could apply our results to probe subleading effects at large but finite time.

\acknowledgments
Work of SCH is supported by the National Science and Engineering Council of Canada, the Canada Research Chair program, the Fonds de Recherche du Qu\'ebec--Nature et Technologies, and the Simons Collaboration on the Nonperturbative Bootstrap.  JS gratefully acknowledges support from the Institut des Sciences Math\'ematiques.
\appendix

\section{Conformal blocks}
\label{app:blocks}

This appendix defines the conformal blocks as used in this paper.
For the fishnet theory we require the $d=4$ blocks, which are given explicitly as
\be \label{G4d}
G^{(a,b)}_{\Delta,J}(z,\zb) = \frac{z\zb}{\zb-z}\big[
 k^{(a,b)}_{\Delta-J-2}(z)k^{(a,b)}_{\Delta+J}(\zb)-(z{\leftrightarrow}\zb)\big]  \qquad\mbox{($d=4$)}\,,
\ee
where
\be
\label{k} k^{(a,b)}_{\beta}(z) = z^{\beta/2}\,{}_2F_1\big(\tfrac{\beta}{2}+a,\tfrac{\beta}{2}+b,\beta,z\big)
\ee
is an eigenfunction of the SL(2,R) Casimir.
In general, conformal blocks are eigenfunctions of the conformal Casimir which
we normalize so that $z,\zb\to 0$ limit contains the following term with unit coefficient:
\be
\lim\limits_{z\ll\zb\ll 1} G^{(a,b)}_{\Delta,J}(z,\zb) = z^{\frac{\Delta-J}{2}}\zb^{\frac{\Delta+J}{2}}\,.
\label{limG}
\ee
This term then comes with an infinite tower of integer powers of $z$, $\zb$.
When the spin is non-integer, this tower is supplemented with a second one such that the combined limit
$z,\zb\to 0$ is a function $C$ \emph{proportional} to the Gegenbauer function $C^{(d/2-1)}_J(x)$:
\be
\lim\limits_{z,\zb\to 0}G^{(a,b)}_{\Delta,J}(z,\zb)  = (z\zb)^{\frac{\Delta}{2}} \P_J\left(\frac{z+\zb}{2\sqrt{z\zb}}\right),
\ee
where
\be \label{def CJ}
 \P_J(x) = \frac{\Gamma\big(\tfrac{d-2}{2}\big)\Gamma(J+d-2)}{\Gamma(d-2)\Gamma\big(J+\tfrac{d-2}{2}\big)}
\  {}_2F_1\big(-J,J+d-2,\tfrac{d-1}{2},\tfrac{1-x}{2}\big).
\ee
The normalization was chosen to be compatible with eq.~\eqref{limG}.

\section{Harmonic polylogarithms and their analytic continuations\label{app:HPL}}
%\subsection{General properties \label{HPLprops}}

Harmonic polylogarithms (HPLs) are generalized logarithms defined and indexed such that
\begin{equation}
	H_1(z)=\int_0^z\frac{d\tilde{z}}{1- \tilde{z}}=-\log(1-z)
\end{equation}
and
\begin{equation}\label{log}
	H_0(z)=\int_0^z\frac{d\tilde{z}}{\tilde{z}}=\log(z).
\end{equation}
Higher-weight HPLs are nested integrals with a binary vector labelling the differential form such that $0$ indicates $d\tilde{z}/\tilde{z}$ and $1$ indicates $d\tilde{z}/(1-\tilde{z})$, as illustrated in the following example \cite{Maitre_2006}:
\begin{equation}\label{integral}
	H_{0,1,1}(z)=\int_0^z \frac{dz'}{z'}\int_0^{z'} \frac{dz''}{1-z''}\int_0^{z''} \frac{dz'''}{1-z'''}.
\end{equation}
It is important to note that HPL functions with a 0 as the rightmost index are singular as $\log(z)^n$ at $z=0$ and those with a 1 at the leftmost index are singular as $\log(1-z)^n$ at $z=1$. This will be a crucial fact in the subsequent section. These divergences can be extracted by considering the integral representation and expanding according to
\begin{equation}\label{logextraction}
	H_{a_1,\ldots,a_k}(z)H_0(z)= H_{a_1,\ldots,a_k,0}(z) + H_{a_1,\ldots,a_{k-1},0,a_k}(z)+\ldots +H_{0,a_1,\ldots,a_k}(z).
\end{equation}
Then, by solving for $H_{a_1,\ldots,a_k,0}(z)$ and recalling equation (\ref{log}), the divergent logarithm is evident. A similar technique allows for the extraction of the $\log(1-z)$ terms.

It is also interesting to note the relation between HPLs and zeta values, namely that $H_{0,\ldots,0,1}(1)=\zeta(n)$, where $n$ is the number of 0 indices and $\zeta$ denotes the Riemann zeta function. In a generalization of this relation, the multiple zeta values (MZV) are defined as
\begin{equation}
	\MZV_{a_1,\ldots a_k} = H_{a_1,\ldots a_k}(1).
\end{equation}

For the correlation function of the conformal fishnet theory in equation (\ref{expansion}), the analytic continuation of $z$ counterclockwise around $z=1$ amounts to extracting the $\log(1-z)$ terms from the HPL functions and replacing them with the additional contribution from moving around the branch cut, $\log(1-z)\to 2\pi i + \log(1-z)$.

\begin{figure}[t]
	\def\svgwidth{0.6\linewidth}
	\centering{
		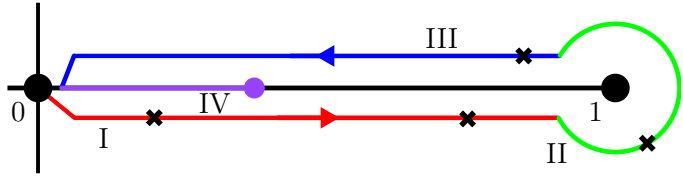
		\caption{Contour of integration in the complex $z$-plane for the calculation of the $C_{\{a_i\}}$ terms. The \textbf{x}'s denote the endpoint $z$ position of the integral $\int_0^z$. Due to the nesting, these $x$'s are ordered. The integral is evaluated by dividing the contour into parts I, II and III. The remaining section, IV, is the original HPL.}
		\label{fig:sw1}
	}
\end{figure}

Alternatively, we can compute the analytic continuation of the HPLs without extracting the logarithms by directly evaluating their analytic continuation. From the integral form of the HPL functions (\ref{integral}), their analytic continuation can be decomposed as
\begin{equation}
	H'_{a_1,\ldots,a_k}(z) =H_{a_1,\ldots,a_k}(z)+ C_{a_1}H_{a_2,\ldots,a_k}(z) + \ldots+
	C_{a_1,\ldots,a_{k-1}}H_{a_k}(z) + C_{a_1,\ldots,a_k},
\end{equation}
where the $C_{\{a_i\}}$'s denote constant contour integrals starting at the $z$-plane origin and looping counterclockwise around $z=1$. These are illustrated in figure \ref{fig:sw1} and can be decomposed into integrals from $z=0\to z=1$, around a countour at $z=1$ and then from $z=1\to z=0$. For example,
\begin{equation}
	C_{1,0,1} = (-1)^3\overline{\MZV}_{1,0,1} - 2\pi i\overline{\MZV}_{1,0} - 2\pi i \MZV_{0,1} +
	\MZV_{1,0,1},
\end{equation}where $\overline{\MZV}$ denotes the MZV value with the binary indices flipped (for example $\overline{\MZV}_{1,0,1}=\MZV_{0,1,0}$). The additional $(-1)^3$ term comes from a change of variables in the integral and the $2\pi i$ comes from the countour integral. In general, these $C_{\{a_i\}}$ constants are calculated by summing all divisions the HPL integrals into the three regions of integration. The middle contour region returns $(-2\pi i)^n/n!$ if all the indices are 1's and zero otherwise. The example above was calculated by considering the integration regions of $C_{1,0,1}$,
\begin{multline}
	C_{1,0,1} = C_{101||} + C_{10|1|} +\color{red} C_{1|01|} \color{black} +\color{blue} C_{10||1}\color{black}+ \color{red}C_{|101|}\color{black} +\color{red}C_{1|0|1}\color{black}+\color{blue}  C_{1||01}\color{black}\\+\color{red}C_{|10|1}\color{black}+C_{|1|01} + C_{||101},
\end{multline}where the $|$'s denote these separated regions, the red indicates terms that vanish and the blue indicating terms that cancel against each other. For example,
\begin{equation}
	C_{|1|01}=|\oint \frac{dz'}{1-z'}|\int_{[0,1]}\frac{dz}{z} \int_{[0,z]} \frac{dz''}{1-z''} = \frac{(-2\pi i)^1}{1!} \MZV_{0,1}.
\end{equation} The $|$ in the equation above symbolically denotes the dividing of the integral rather than an absolute value.

By analytically continuing the HPL functions of $z$ in the fishnet correlator expansion, this continuation technique provides equivalent results to the $\log(1-z)$ replacement while drastically reducing computation time.

\section{Froissart-Gribov formula and cancellation of spurious poles}
\label{app:FG}

In section \ref{confreggesmatrix}, we claimed that all the spurious poles cancel against each other in the Sommerfeld-Watson resummation for flat space scattering. In this appendix, we provide justification.
To understand the necessary cancellations, we need a concrete expression
for the coefficients $a^{t,u}_J$ known as the Froissart-Gribov formula.
In brief (this is reviewed in \cite{Collins:1977jy,Donnachie:2002en}, see also \cite{Caron-Huot:2017vep,Correia:2020xtr})
we may use the orthogonality of spherical harmonics to write partial wave coefficients (for integer $J$)
as an integral over $[-1,1]$ against the polynomial solution $\P_J$,
which is equal to an integral over the discontinuity of the nonpolynomial solution $\Ppure$ introduced in eq.~\eqref{P from Ppures}.
This allows the contour to be deformed to pick the discontinuities of $F(x)$:
\bea
 a_J&=& \frac{\Gamma(J+\tfrac{d}{2})}{\sqrt{\pi}\Gamma(\tfrac{d-1}{2})\Gamma(J+1)}
 \int_{-1}^1 dx (1-x^2)^{\frac{d-3}{2}} F(x) \tilde\P_{J}(x) \label{aJ Euclidean}\\
 &=& \frac{1}{i\pi} \oint_{[-1,1]} dx \left(\frac{x^2-1}{4}\right)^{\frac{d-3}{2}}F(x) \Ppure_{2-d-J}(x) \\
\Rightarrow\quad
 a_J^t&=&
 \int_{x_0}^\infty \left(\frac{x^2-1}{4}\right)^{\frac{d-3}{2}} \Ppure_{2-d-J}(x)  \frac{F(x+i0)-F(x-i0)}{i\pi} 
 \label{FG}
\eea
where $\tilde\P_{J}(x)$ is just the hypergeometric function in eq.~\eqref{def CJ} (without $\Gamma$-factors),
the contour on the second line encircles the cut of $\Ppure_{2-d-J}(x)$ counter-clockwise,
and on the third line we assumed that singularities of $F(x)$ consist of a right-cut for $x>x_0$
(and a left-cut for $x<-x_0$ which give $a_J^u$ in the decomposition \eqref{atu}).
A technical comment: the Froissart-Gribov integral (\ref{FG}) is valid for $J$ large enough that we can ignore arcs at infinity: ${\rm Re}(J)>j_*$ if $F\sim x^{j*}$. To the left of that, the analytic continuation of $a_J^t$ need not agree with the coefficients entering eq.~\eqref{flat OPE}, whence the subtraction terms in eq.~\eqref{flat WS}.

We can now explain the two mechanisms responsible for spurious pole cancellation.

We begin with the cancellation of type-2 and type-3 poles defined above eq.~\eqref{Regge}.
The Froissart-Gribov integral \eqref{FG} produces singularities for two reasons:
``physical'' singularities from divergences of the integral, and ``spurious'' poles due to the integrand itself.
It is helpful to denote the two solutions of the Gegenbauer equation, for a given value of $J$, as ``small'' and ``large'' depending on whether they vanish or grow as $x\to\infty$.
Large solutions can have poles with residue proportional to the small solution.
In the right half-plane the small solution is $\Ppure_{2-d-J}$ but the roles get exchanged when ${\rm Re}(J)=-\frac{d-2}{2}$,
and so it acquires the following spurious left pole:
\be
\Res\limits_{J=-\tfrac{d-2}{2}-m}\Ppure_{2-d-J}(x)\propto
%= \frac{(-1)^{m+1}\big(\tfrac{4-d}{2}-m\big)_{2m}}{(m-1)!m!}
\Ppure_{-\tfrac{d-2}{2}-m}(x)\qquad\mbox{($m=1,2,3\ldots$)}\,. \label{spurious prop}
\ee
These are the spurious left-poles of $a_J$ called type-2 above eq.~\eqref{Regge}.
On the other hand, the right-poles of type-3 come from the combination $(P_J(x) - \Ppure_J(x))$ for which we deform the contour
to the right in deriving eq.~\eqref{Regge}.
Since $P_J$ is pole-free on the right, this combination has the same spurious pole as the large solution $-\Ppure_J(x)$.  Thus in the contour deformation argument leading to eq.~\eqref{Regge} we see that all spurious poles come from the following
which is always a product of small and large solution:
\be \label{spurious 23}
\frac{e^{-i\pi J} a_J^t +a_J^u}{\sin(\pi J)}\Ppure_J(x) \to \left\{\begin{array}{l}
\mbox{``large''$\times$``small''}, \quad \mbox{left half-plane},\\
\mbox{``small''$\times$``large''}, \quad \mbox{right half-plane},
\end{array}\right.
\ee
and it is easy to see that the residue of both left- and right- spurious poles is the product of the same two small solutions
(one hidden in $a_J$ and the explicit function of $x$), allowing them to precisely cancel.
It is interesting to check directly that the proportionality constant in eq.~\eqref{spurious prop} is precisely
the residue of gamma factors in eq.~\eqref{P from Ppures}, as guaranteed by the general argument just given.

We stress that we are not excluding the possibility that $a_J$ may contain a physical pole at
$J={-}\tfrac{d-2}{2}{-}m$ (coming from a large-$x$ divergence in the Froissart-Gribov integral).
Rather, the message is that in such a case the ``physical'' residue to be included in the Regge formula \eqref{Regge} is the residue of $a_J$ minus a spurious part proportional to $a^t_{-\frac{d-2}{2}+m}$ (similar to the way $\Delta$-plane
spurious poles are treated in eq.~(3.9) of \cite{Caron-Huot:2017vep}).

The second mechanism, responsible for cancellation of type-4 poles, is the identity:
\be
 a^t_J + (-1)^J a^u_J  = 0 \qquad \mbox{(for $J=-1,-2,-3\ldots$)}
\ee
valid in generic theories which follows from the explicit $1/\Gamma(J+1)$ in eq.~(\ref{aJ Euclidean}).
By generic we mean that
the combination $a_J/\sin(\pi J)$ has a pole at negative integer $J$ if only if the Froissart-Gribov formula exhibits
a large-$x$ divergence at that exponent, which is non-generic.
We note that these mechanisms are only distinct in generic spacetime dimension: for SO($d$) partial waves with $d$ even,
type-3 and type-4 poles occur at the same location. This case is discussed in appendix A of \cite{Donnachie:2002en}.
The overall cancellations (and the result \eqref{Regge}) remain valid but one has to use both mechanism at the same time.

% Exp[-I Pi j]
%1/(Exp[-I pi j]-Exp[I pi j])
\section{Lorentzian inversion of Regge block in $d=2$ and $d=4$}
\label{app:Plancherel}

In this appending we compute the pairing  defined in eq.~\eqref{pairing} between ``funny block'' and Regge block,
which we recall here for convenience:
\be
\frac{\kappa_{\Delta+J}^{(a,b)}}{4}
\int_0^1 dzd\zb\ \mu(z,\zb)\ G^{(-a,-b)}_{J+d-1,\Delta+1-d}(z,\zb) \frac{R^{(a,b)}_{\Delta,J}(z,\zb)}{2\kappa_{\Delta'+J'}^{(a,b)}}
\equiv P^{(a,b)}_{\Delta,J; \Delta',J'}\,. \label{pairing app}
\ee
For technical reasons, we were only able to compute this pairing in $d=2$ and $d=4$ spacetime dimensions.
Two-dimensional blocks admit the following simple form:
\be
G^{(a,b)}_{\Delta,J}(z,\zb) = \frac{k^{(a,b)}_{\Delta-J}(z)k^{(a,b)}_{\Delta+J}(\zb)+(z{\leftrightarrow}\zb)}{1+\delta_{J,0}}
\qquad (d=2) \label{G2d}
\ee
where $k^{(a,b)}_{\Delta-J}(z)$ stand for hypergeometric functions defined in eq.~\eqref{k}.
They are eigenfunctions of the SL(2,R) Casimir, which we will need shortly:
\be \label{casimir}
 \left( z^2\partial_z (1-z)\partial_z -(a+b)z^2\partial_z -a b\right) k^{(a,b)}_{\beta}(z) = \tfrac14\beta(\beta-2)k^{(a,b)}_{\beta}(z)\,.
\ee
From the defining property \eqref{disc R} it is easy to see that
the Regge block must admit a factorized expression similar to eq.~\eqref{G2d}, with $k$ replaced by
involving $k^\prime$, which is the solution to same equation but whose discontinuity around $z=1$ vanishes:
\ba
k^{\prime (a,b)}_{\beta}(z) &= k^{(a,b)}_{\beta}(z) - \kappat^{(a,b)}_{2-\beta}k^{(a,b)}_{2-\beta}(z)
\\ &= \frac{\pi}{\Gamma(1+a+b)}\frac{r^{(-a,-b)}_{\beta}}{s^{(-a,-b)}_{\beta}}\ z^{\frac\beta2}{}_2F_1\big(\tfrac{\beta}{2}+a,\tfrac{\beta}{2}+b,1+a+b,1-z\big)\,.
\ea
Evaluating the definition \eqref{Rab} in $d=2$ we find exactly this!
\be \label{kp}
R^{(a,b)}_{\Delta,J} = k^{\prime(a,b)}_{\Delta-J}(z)k^{\prime(a,b)}_{2-\Delta-J}(\zb)+(z{\leftrightarrow}\zb)\,.
\ee
A similar substitution works for the $d=4$ blocks in eq.~\eqref{G4d}.
The pairing then reduces to the following one-dimensional pairing between $k$ and $k^\prime$ blocks:
\be
 P^{(a,b)1d}_{\beta,\beta'} \equiv \int_0^1 \frac{dz}{z^2} k^{(-a,-b)}_\beta(z) k^{\prime(a,b)}_{\beta'}(z)\,.
\ee
To compute this, we use the Casimir equation (\ref{casimir}) and integrate by parts, which gives
\ba \label{casimir trick}
\tfrac14 (\beta(\beta-2)-\beta'(\beta'-2))P^{(a,b)1d} &= \lim_{z\to 1} (1-z)
\left[k^{(-a,-b)}_\beta(z) \left(\overset{\leftarrow}{\partial}_z-\overset{\rightarrow}{\partial}_z+\frac{a+b}{1-z}\right)k^{\prime(a,b)}_{\beta'}(z)\right] \\
&= (1-\beta')\frac{r_{\beta}^{(a,b)}}{r_{2-\beta'}^{(a,b)}}\,.
\ea
The result is simple because only the regular part of $k^{(-a,-b)}_\beta(z)$ contributes to the limit.
Substituting this into eq.~\eqref{pairing app} for both $d=2$ and $d=4$,
and we find a common, compelling expression for both:
\ba
P^{(a,b)}_{\Delta,J; \Delta',J'}&=
 \frac{r_{J+d-\Delta}^{(a,b)} r_{J'+\Delta'}^{(-a,-b)}}{r_{J'+d-\Delta'}^{(a,b)}r_{J+\Delta}^{(-a,-b)}}
 \frac{4(J+\Delta-1) (J'+ d - \Delta' - 1)}{(\Delta{-}\Delta'{+}J{-}J')(\Delta{-}\Delta'{-}J{+}J')(\Delta{-}\tilde{\Delta}'{-}J{+}\tilde{J'})
(\Delta{-}\tilde{\Delta}'{+}J{-}\tilde{J'})}\\
&\phantom{=} + \mbox{($\Delta'$ shadow)}\,.
\label{pairing answer}
\ea
This formula is further discussed in the main text, see eq.~\eqref{pairing answer text}, where it is
interpreted as a Plancherel formula for the SO($d$,2) group.
The common denominator of the two combined lines suggests that the Casimir trick
used in 1d could be extended to general spacetime dimensions using a eighth order differential operator,
but we found this idea somewhat challenging to implement.

\bibliography{refs}

\providecommand{\href}[2]{#2}\begingroup\raggedright\begin{thebibliography}{10}

\bibitem{Brower:2006ea}
R.~C. Brower, J.~Polchinski, M.~J. Strassler, and C.-I. Tan, ``{The Pomeron and
  gauge/string duality},''
  \href{http://dx.doi.org/10.1088/1126-6708/2007/12/005}{{\em JHEP} {\bfseries
  12} (2007) 005}, \href{http://arxiv.org/abs/hep-th/0603115}{{\ttfamily
  arXiv:hep-th/0603115}}.

\bibitem{Cornalba:2007fs}
L.~Cornalba, ``{Eikonal methods in AdS/CFT: Regge theory and multi-reggeon
  exchange},'' \href{http://arxiv.org/abs/0710.5480}{{\ttfamily arXiv:0710.5480
  [hep-th]}}.

\bibitem{Costa:2012cb}
M.~S. Costa, V.~Goncalves, and J.~Penedones, ``{Conformal Regge theory},''
  \href{http://dx.doi.org/10.1007/JHEP12(2012)091}{{\em JHEP} {\bfseries 12}
  (2012) 091},
\href{http://arxiv.org/abs/1209.4355}{{\ttfamily arXiv:1209.4355 [hep-th]}}.
%%CITATION = ARXIV:1209.4355;%%.

\bibitem{Maldacena:2015waa}
J.~Maldacena, S.~H. Shenker, and D.~Stanford, ``{A bound on chaos},''
  \href{http://dx.doi.org/10.1007/JHEP08(2016)106}{{\em JHEP} {\bfseries 08}
  (2016) 106}, \href{http://arxiv.org/abs/1503.01409}{{\ttfamily
  arXiv:1503.01409 [hep-th]}}.

\bibitem{Camanho:2014apa}
X.~O. Camanho, J.~D. Edelstein, J.~Maldacena, and A.~Zhiboedov, ``{Causality
  Constraints on Corrections to the Graviton Three-Point Coupling},''
  \href{http://dx.doi.org/10.1007/JHEP02(2016)020}{{\em JHEP} {\bfseries 02}
  (2016) 020}, \href{http://arxiv.org/abs/1407.5597}{{\ttfamily arXiv:1407.5597
  [hep-th]}}.

\bibitem{Hartman:2016lgu}
T.~Hartman, S.~Kundu, and A.~Tajdini, ``{Averaged Null Energy Condition from
  Causality},'' \href{http://dx.doi.org/10.1007/JHEP07(2017)066}{{\em JHEP}
  {\bfseries 07} (2017) 066}, \href{http://arxiv.org/abs/1610.05308}{{\ttfamily
  arXiv:1610.05308 [hep-th]}}.

\bibitem{Liu:2020tpf}
J.~Liu, D.~Meltzer, D.~Poland, and D.~Simmons-Duffin, ``{The Lorentzian
  inversion formula and the spectrum of the 3d O(2) CFT},''
  \href{http://arxiv.org/abs/2007.07914}{{\ttfamily arXiv:2007.07914
  [hep-th]}}.

\bibitem{Caron-Huot:2020ouj}
S.~Caron-Huot, Y.~Gobeil, and Z.~Zahraee, ``{The leading trajectory in the 2+1D
  Ising CFT},'' \href{http://arxiv.org/abs/2007.11647}{{\ttfamily
  arXiv:2007.11647 [hep-th]}}.

\bibitem{Cornalba:2007zb}
L.~Cornalba, M.~S. Costa, and J.~Penedones, ``{Eikonal approximation in
  AdS/CFT: Resumming the gravitational loop expansion},''
  \href{http://dx.doi.org/10.1088/1126-6708/2007/09/037}{{\em JHEP} {\bfseries
  09} (2007) 037}, \href{http://arxiv.org/abs/0707.0120}{{\ttfamily
  arXiv:0707.0120 [hep-th]}}.

\bibitem{Li:2017lmh}
D.~Li, D.~Meltzer, and D.~Poland, ``{Conformal Bootstrap in the Regge Limit},''
  \href{http://dx.doi.org/10.1007/JHEP12(2017)013}{{\em JHEP} {\bfseries 12}
  (2017) 013},
\href{http://arxiv.org/abs/1705.03453}{{\ttfamily arXiv:1705.03453 [hep-th]}}.
%%CITATION = ARXIV:1705.03453;%%.

\bibitem{Costa:2017twz}
M.~S. Costa, T.~Hansen, and J.~Penedones, ``{Bounds for OPE coefficients on the
  Regge trajectory},'' \href{http://dx.doi.org/10.1007/JHEP10(2017)197}{{\em
  JHEP} {\bfseries 10} (2017) 197},
\href{http://arxiv.org/abs/1707.07689}{{\ttfamily arXiv:1707.07689 [hep-th]}}.
%%CITATION = ARXIV:1707.07689;%%.

\bibitem{Kravchuk:2020scc}
P.~Kravchuk, J.~Qiao, and S.~Rychkov, ``{Distributions in CFT. Part I.
  Cross-ratio space},'' \href{http://dx.doi.org/10.1007/JHEP05(2020)137}{{\em
  JHEP} {\bfseries 05} (2020) 137},
  \href{http://arxiv.org/abs/2001.08778}{{\ttfamily arXiv:2001.08778
  [hep-th]}}.

\bibitem{Gurdogan:2015csr}
O.~Gurdogan and V.~Kazakov, ``{New Integrable 4D Quantum Field Theories from
  Strongly Deformed Planar $\mathcal N = $ 4 Supersymmetric Yang-Mills
  Theory},'' \href{http://dx.doi.org/10.1103/PhysRevLett.117.201602}{{\em Phys.
  Rev. Lett.} {\bfseries 117} no.~20, (2016) 201602},
  \href{http://arxiv.org/abs/1512.06704}{{\ttfamily arXiv:1512.06704
  [hep-th]}}. [Addendum: Phys.Rev.Lett. 117, 259903 (2016)].

\bibitem{Gromov:2018hut}
N.~Gromov, V.~Kazakov, and G.~Korchemsky, ``{Exact Correlation Functions in
  Conformal Fishnet Theory},''
  \href{http://dx.doi.org/10.1007/JHEP08(2019)123}{{\em JHEP} {\bfseries 08}
  (2019) 123},
\href{http://arxiv.org/abs/1808.02688}{{\ttfamily arXiv:1808.02688 [hep-th]}}.
%%CITATION = ARXIV:1808.02688;%%.

\bibitem{Korchemsky:2018hnb}
G.~Korchemsky, ``{Exact scattering amplitudes in conformal fishnet theory},''
  \href{http://dx.doi.org/10.1007/JHEP08(2019)028}{{\em JHEP} {\bfseries 08}
  (2019) 028}, \href{http://arxiv.org/abs/1812.06997}{{\ttfamily
  arXiv:1812.06997 [hep-th]}}.

\bibitem{Chowdhury:2019hns}
S.~Dutta~Chowdhury, P.~Haldar, and K.~Sen, ``{On the Regge limit of Fishnet
  correlators},'' \href{http://dx.doi.org/10.1007/JHEP10(2019)249}{{\em JHEP}
  {\bfseries 10} (2019) 249}, \href{http://arxiv.org/abs/1908.01123}{{\ttfamily
  arXiv:1908.01123 [hep-th]}}.

\bibitem{Chowdhury:2020tbn}
S.~D. Chowdhury, P.~Haldar, and K.~Sen, ``{Regge amplitudes in Generalized
  Fishnet and Chiral Fishnet Theories},''
  \href{http://arxiv.org/abs/2008.10201}{{\ttfamily arXiv:2008.10201
  [hep-th]}}.

\bibitem{Collins:1977jy}
P.~D.~B. Collins, {\em {An Introduction to Regge Theory and High-Energy
  Physics}}.
\newblock Cambridge Monographs on Mathematical Physics. Cambridge Univ. Press,
  Cambridge, UK, 2009.
\newblock
\url{http://www-spires.fnal.gov/spires/find/books/www?cl=QC793.3.R4C695}.
\newblock
%%CITATION = INSPIRE-127083;%%.

\bibitem{Donnachie:2002en}
S.~Donnachie, H.~G. Dosch, O.~Nachtmann, and P.~Landshoff, ``{Pomeron physics
  and QCD},''
{\em Camb. Monogr. Part. Phys. Nucl. Phys. Cosmol.} {\bfseries 19} (2002)
  1--347.
%%CITATION = CMPCE,19,1;%%.

\bibitem{Simmons-Duffin:2017nub}
D.~Simmons-Duffin, D.~Stanford, and E.~Witten, ``{A spacetime derivation of the
  Lorentzian OPE inversion formula},''
  \href{http://dx.doi.org/10.1007/JHEP07(2018)085}{{\em JHEP} {\bfseries 07}
  (2018) 085},
\href{http://arxiv.org/abs/1711.03816}{{\ttfamily arXiv:1711.03816 [hep-th]}}.
%%CITATION = ARXIV:1711.03816;%%.

\bibitem{SimmonsDuffin:2012uy}
D.~Simmons-Duffin, ``{Projectors, Shadows, and Conformal Blocks},''
  \href{http://dx.doi.org/10.1007/JHEP04(2014)146}{{\em JHEP} {\bfseries 04}
  (2014) 146}, \href{http://arxiv.org/abs/1204.3894}{{\ttfamily arXiv:1204.3894
  [hep-th]}}.

\bibitem{Caron-Huot:2017vep}
S.~Caron-Huot, ``{Analyticity in Spin in Conformal Theories},''
  \href{http://dx.doi.org/10.1007/JHEP09(2017)078}{{\em JHEP} {\bfseries 09}
  (2017) 078},
\href{http://arxiv.org/abs/1703.00278}{{\ttfamily arXiv:1703.00278 [hep-th]}}.
%%CITATION = ARXIV:1703.00278;%%.

\bibitem{Kravchuk:2018htv}
P.~Kravchuk and D.~Simmons-Duffin, ``{Light-ray operators in conformal field
  theory},'' \href{http://dx.doi.org/10.1007/JHEP11(2018)102}{{\em JHEP}
  {\bfseries 11} (2018) 102}, \href{http://arxiv.org/abs/1805.00098}{{\ttfamily
  arXiv:1805.00098 [hep-th]}}.

\bibitem{Isachenkov:2017qgn}
M.~Isachenkov and V.~Schomerus, ``{Integrability of conformal blocks. Part I.
  Calogero-Sutherland scattering theory},''
  \href{http://dx.doi.org/10.1007/JHEP07(2018)180}{{\em JHEP} {\bfseries 07}
  (2018) 180}, \href{http://arxiv.org/abs/1711.06609}{{\ttfamily
  arXiv:1711.06609 [hep-th]}}.

\bibitem{Raben:2018sjl}
T.~G. Raben and C.-I. Tan, ``{Minkowski conformal blocks and the Regge limit
  for Sachdev-Ye-Kitaev-like models},''
  \href{http://dx.doi.org/10.1103/PhysRevD.98.086009}{{\em Phys. Rev. D}
  {\bfseries 98} no.~8, (2018) 086009},
  \href{http://arxiv.org/abs/1801.04208}{{\ttfamily arXiv:1801.04208
  [hep-th]}}.

\bibitem{Kos:2013tga}
F.~Kos, D.~Poland, and D.~Simmons-Duffin, ``{Bootstrapping the $O(N)$ vector
  models},'' \href{http://dx.doi.org/10.1007/JHEP06(2014)091}{{\em JHEP}
  {\bfseries 06} (2014) 091}, \href{http://arxiv.org/abs/1307.6856}{{\ttfamily
  arXiv:1307.6856 [hep-th]}}.

\bibitem{Maitre_2006}
D.~Maitre, ``Hpl, a mathematica implementation of the harmonic
  polylogarithms,'' \href{http://dx.doi.org/10.1016/j.cpc.2005.10.008}{{\em
  Computer Physics Communications} {\bfseries 174} no.~3, (Feb, 2006)
  222--240}. \url{http://dx.doi.org/10.1016/j.cpc.2005.10.008}.

\bibitem{Correia:2020xtr}
M.~Correia, A.~Sever, and A.~Zhiboedov, ``{An Analytical Toolkit for the
  S-matrix Bootstrap},'' \href{http://arxiv.org/abs/2006.08221}{{\ttfamily
  arXiv:2006.08221 [hep-th]}}.

\end{thebibliography}\endgroup
\bibliographystyle{utphys}

\end{document}